\documentclass[journal,comsoc]{IEEEtran}

\usepackage{hyperref}
\usepackage{graphicx}
\usepackage{epsfig}
\usepackage{algorithm}
\usepackage{algorithmic}
\usepackage{ifmtarg}
\usepackage[labelsep=none]{caption}
\usepackage{footnote}
\usepackage{cite}
\usepackage{amssymb}
\usepackage{amsthm}
\usepackage{amsmath}
\usepackage{cleveref}
\usepackage{color}
\usepackage{caption}
\usepackage{subcaption}
\usepackage{epstopdf}

\begin{document}

\title{Green Networking in Cellular HetNets: A Unified Radio Resource Management Framework with Base Station ON/OFF Switching}
\author{
\IEEEauthorblockN{\large Hakim Ghazzai, \textit{Member, IEEE}, Muhammad Junaid Farooq, \textit{Student Member, IEEE}, Ahmad Alsharoa, \textit{Student Member, IEEE}, Elias Yaacoub, \textit{Senior Member, IEEE}, Abdullah Kadri, \textit{Senior Member, IEEE} and Mohamed-Slim Alouini, \textit{Fellow, IEEE}\\}\vspace{-0.5cm}
\thanks{ \hrule
\vspace{0.1cm} \indent Copyright (c) 2015 IEEE. Personal use of this material is permitted. However, permission to use this material for any other purposes must be obtained from the IEEE by sending a request to pubs-permissions@ieee.org.

A part of this work has been published in IEEE Vehicular Technology Conference (VTC-spring 2015), Glasgow, Scotland, UK.

This work was made possible, in part, by NPRP grant \# 6-001-2-001 from the Qatar National Research Fund (A member of The Qatar Foundation). The statements made herein are solely the responsibility of the authors.

Hakim Ghazzai, Muhammad Junaid Farooq, and Abdullah Kadri are with Qatar Mobility Innovations Center (QMIC), Qatar University, Doha, Qatar. E-mails: \{junaidf, hakimg, abdullahk\}@qmic.com.

Ahmad Alsharoa is with Electrical and Computer Engineering Department, Iowa State University, Ames, Iowa, USA. E-mail: alsharoa@iastate.edu.

Elias Yaacoub is with the Faculty of Computer Studies, Arab Open University (AOU), Beirut, Lebanon. E-mail: eliasy@ieee.org.

Mohamed-Slim Alouini is with King Abdullah University of Science and Technology (KAUST), Thuwal, Makkah Province, Saudi Arabia. E-mail: slim.alouini@kaust.edu.sa.

This work was done while Muhammad Junaid Farooq was at QMIC. He is now in New York University, NY, United States.
}}

\maketitle
\thispagestyle{empty}
\pagestyle{empty}

\begin{abstract}
\boldmath{In this paper, the problem of energy efficiency in cellular heterogeneous networks (HetNets) is investigated using radio resource and power management combined with the base station (BS) ON/OFF switching. The objective is to minimize the total power consumption of the network while satisfying the quality of service (QoS) requirements of each connected user. We consider the case of co-existing macrocell BS, small cell BSs, and private femtocell access points (FAPs). Three different network scenarios are investigated, depending on the status of the FAPs, i.e., HetNets without FAPs, HetNets with closed FAPs, and HetNets with semi-closed FAPs. A unified framework is proposed to simultaneously allocate spectrum resources to users in an energy efficient manner and switch off redundant small cell BSs. The high complexity dual decomposition technique is employed to achieve optimal solutions for the problem. A low complexity iterative algorithm is also proposed and its performances are compared to those of the optimal technique. The particularly interesting case of semi-closed FAPs, in which the FAPs accept to serve external users, achieves the highest energy efficiency due to increased degrees of freedom. In this paper, a cooperation scheme between FAPs and mobile operator is also investigated. The incentives for FAPs, e.g., renewable energy sharing and roaming prices, enabling cooperation are discussed to be considered as a useful guideline for inter-operator agreements.}
\end{abstract}

\begin{IEEEkeywords}
Base station ON/OFF switching, femtocell access points, green networking, heterogeneous networks.
\end{IEEEkeywords}

\section{Introduction}\label{Introduction}
With the ever increasing number of mobile broadband data users and bandwidth-intensive services, the demand for radio resources has increased tremendously. One of the methods used by mobile operators to meet this challenge is to deploy additional low-powered base stations (BSs) in areas of high demand. The resulting network, referred to as heterogeneous network (HetNet)~\cite{HetNet}, helps in maintaining the quality of service (QoS) for a larger number of users by reusing the spectrum. However, with the densification of these HetNets, energy consumption and hence the carbon footprint have significantly raised. Therefore, conserving energy while meeting the user's QoS requirements has been the focus of the green communications researchers.

Most of the wireless data usage is in indoor environments such as offices, residential buildings, shopping malls, etc., where the users may face difficulties in achieving high data rates while connecting to the macrocell BSs. This is mainly due to the penetration loss incurred by the wireless signals inside the buildings. Therefore, to increase the capacity of the network in these hotspots, small cell BSs\footnote{The term small cell BS in this paper refers to picocell BS, which is much smaller than macrocell BS and larger than femtocell BS (typically has a range of 100 or 300 meters).} are deployed in close proximity to the buildings~\cite{small_cell_1,int_1}. In addition to small cell BSs, there are often additional privately owned femtocell access points (FAPs) or home BSs installed inside buildings that co-exist alongside macrocell BS and small cell BSs~\cite{3GPPTS22220}. These FAPs use the local broadband connection to provide cellular services to their subscribers.

Small cell BSs provide increased coverage and network capacity during peak times, however, they might not be very useful under light traffic load scenarios. Instead, they might be under-utilized or completely redundant leading to inefficient use of energy and communication resources. Hence, during periods of low traffic, it is appropriate to turn off the small cell BSs and offload the users to a nearby macrocell BS. The presence of FAPs provides additional flexibility in this regard, since the small cell users can be offloaded to the FAPs, if it is permitted. In fact, FAP access control schemes can be classified into three categories~\cite{femto_capacity,sleep1}, i.e., open access, closed access, and hybrid or semi-closed access. Open access allows a nearby mobile subscriber to connect to the FAP without any restriction, thus acting as a small cell BS. On the other hand, closed access is only limited to registered subscribers, also called femto owners. In hybrid access, unregistered users connected to the macrocell or small cell BS might be allowed to use some FAP resources subject to their availability and certain constraints. If private FAPs cooperate with the mobile operator, there will be an increased potential of saving considerable amount of energy without compromising the subscribers' QoS.

Although energy consumption cost and the environmental impact of redundant small cell BSs can be reduced by locally generating renewable energy at the small cell BS sites, cooperation with private FAPs provides an opportunity to further improve energy efficiency. Also, the impact of such energy saving is high since cooperation may allow some small cell BSs, which consume higher energy than FAPs, to be turned off that would otherwise be on. In addition, the mobile operator can use a portion of its energy savings, achieved from cooperation, as well as excess resources, e.g., renewable energy, to provide incentives to the FAP operator.

To exploit this envisioned potential for energy saving, and hence the carbon footprint reduction in cellular HetNets, an effective resource and power management framework is required along with the BS ON/OFF strategy. The proposed framework aims to efficiently allocate spectrum resources to users in a way that cross-tier interference is minimum and hence minimum transmit power is used by the BSs. Moreover, it will decide the optimal set of small cell BSs that should remain active given a pre-defined spatial distribution of users. Finally, it will specify the conditions for FAP cooperation on the basis of which agreements can be reached with the mobile operator.

\subsection{Related Work}
In the recent years, many studies have been proposed to investigate the energy efficiency problem of cellular HetNets using dynamic BS ON/OFF switching technique~\cite{energy_efficiency_1,small_cell_BS_ON_OFF,7389333}. In~\cite{energy_efficiency_1}, the impact of turning off macrocell BSs on the energy efficiency of the HetNet is studied while keeping the small cell BSs active. In contrast,~\cite{small_cell_BS_ON_OFF} investigates the energy saving potential of HetNets using different sleeping strategies for small cell BSs. In~\cite{7389333}, the authors studied the dynamic small-cell ON/OFF operation while serving the offloaded traffic using the macrocell BS to minimize the total power consumption of the HetNet. Two algorithms are proposed to make ON/OFF switching decision. The first one is an optimal location-based operation algorithm applied in the case of uniformly distributed users. The second algorithm is a suboptimal approach proposed for non-uniformly distributed users. It has been shown that power saving is achieved thanks to the proposed dynamic ON/OFF switching operation. In~\cite{sleep1}, three different approaches for small cell BS ON/OFF switching in HetNets are discussed. The ON/OFF status of the small cell BSs is controlled by either the detection of active users by the small cell BSs, wake-up signals by the core network, or wake-up signals by the users. The authors of~\cite{6519623} have introduced two sleep modes to cater for the short and long idle periods of the users. It is shown that dense HetNets can be used to achieve higher capacity and performance while simultaneously reducing energy consumption.

Another dimension of energy saving in HetNets is the management of cross-tier interference~\cite{6171999,int_2}. With reduced interference, the transmit power of BSs can be reduced while achieving the same QoS level. Several approaches are used in literature to manage the interference problem~\cite{energy_efficiency_2,Resource_allocation_3,crosstier}. One of the approaches is the radio resource management that divides the available spectrum into sub-frequency bands where the femtocells and small cell BSs use non-overlapping spectrum bands with the macrocell BS~\cite{energy_efficiency_2}. A decentralized spectrum allocation policy for two-tier (macrocell and femtocell) orthogonal frequency division multiple access (OFDMA) networks is provided in~\cite{Resource_allocation_3}. In~\cite{crosstier}, an iterative algorithm for optimal power and resource block allocation is proposed for two-tier HetNets with a cross-tier interference constraint. Finally, a survey on self-configuring and self-optimizing techniques for interference management in OFDMA HetNets is provided in~\cite{Resource_allocation_1}. A second approach for managing cross-tier interference is related to spectrum sensing, whereby femtocells can sense the activity of the macrocell and opportunistically use the available spectrum~\cite{sensing_1,sensing_2}. The authors in~\cite{sensing_1} proposed a cognitive resource management scheme inspired from the spirit of cognitive radio technology that mitigates the interference between macrocell BSs and small cell BSs. Similarly, in~\cite{sensing_2}, the authors proposed a priority based frequency resource allocation scheme that favors the macrocell over the femtocell. Another method to counter cross-tier interference in HetNets is to use power control. In \cite{power_control_1}, a self-optimized coverage coordination framework between macrocell and femtocell is proposed based on the distributed power control by the femtocell BSs. In~\cite{power_control_2}, a power control approach for two-tier femtocell networks is introduced to reduce the transmit power levels of the strongest femtocell interferers.

In addition to energy conservation methods, the use of renewable energy to power cellular HetNets is also considered in literature~\cite{7570259,systems_hetnets,green_hybrid}. Availability of locally generated renewable energy helps in reducing the carbon footprint of the BSs by limiting the required fossil fuel based energy. For instance in~\cite{systems_hetnets}, a HetNet with hybrid energy supplies, i.e., the small cell BSs are powered by either the traditional electricity grid or green energy sources, has been used to minimize energy costs of the network. In~\cite{green_hybrid}, user association in spatial domain and the renewable energy allocation in the time domain are optimized to reduce the operational expenditures of cellular networks.

\subsection{Contributions}\label{Contributions}
Existing works for energy saving in cellular HetNets consider either BS ON/OFF strategy~\cite{sleep1,switching,6399220,6519623} or energy efficient resource allocation~\cite{energy_efficiency_2,Resource_allocation_1,crosstier,Resource_allocation_3,7145965}. However, the joint resource and transmit power management for cellular HetNets including macrocell BS, small cell BSs and private FAPs along with BS ON/OFF strategy has not been investigated previously. In contrast to literature, we develop a unified framework that aims to reduce energy consumption using efficient resource allocation, transmit power allocation, and BS ON/OFF switching. We also incorporate the availability of renewable energy sources in the proposed framework. Additionally, we investigate the cooperative scenario between cellular HetNets and private femtocell networks to form a virtual network, which contributes significantly in the resource allocation and BS ON/OFF decision.

It should be noted that, in this study, we are developing a green networking technique for HetNets where the objective is to curtail negative environmental impacts of cellular networks by minimizing their total energy consumption and exploiting renewable energy whenever it is possible. Indeed, radio access networks are consuming around 80\% of the total energy consumed by a cellular network~\cite{Nokia}. Moreover, recent studies show that for a typical cellular network, 50\% of the deployed BSs serve only 15$\%$ of the total traffic, while 5\% of the sites serve 20\% of the traffic~\cite{Ericsson}. Therefore, we focus essentially on this part of the network and correspondingly the downlink direction. The main contributions of this paper are summarized as follows:

\begin{itemize}
\item An optimization problem aimed at minimizing the total power consumption of cellular HetNets is formulated. Problem formulation takes into account the power budget of the BSs and the cross-tier interference while respecting QoS requirements, i.e., minimum data rate for each served user. Energy efficiency is achieved using effective radio resource management, power allocation, and BS ON/OFF switching. Different network scenarios emanating from the FAP access method, i.e., open access, closed access, and hybrid access, are considered.
\item The optimal carrier assignment and power allocation solution for the cellular HetNet is reached using the dual decomposition technique. Due to the high computational complexity of the optimal solution, a practical and low complexity algorithm is also developed and the obtained results are compared with those of the optimal solution.
\item A cooperative framework between mobile networks and private FAPs, using hybrid access, is proposed. An economically viable criterion for cooperation agreements is also developed taking into account the interests of both the mobile operator as well as the FAP operators. The incentives from the mobile operator to the FAP operator include cost effective renewable energy sharing and direct monetary payment to cover the additional costs incurred by the FAPs.
\end{itemize}

The remainder of this paper is organized as follows: Section~\ref{System Model} presents the system model. The problem formulation and the different network scenarios are presented in Section~\ref{ProblemFormulation}. The dual decomposition method and the low complexity solution are investigated in~\ref{Solutions}. The cooperation scheme in the context of hybrid access is presented in Section~\ref{CoopAgr}. The numerical results are presented and explained in Section~\ref{Simulation}. Finally, the paper is concluded in Section~\ref{Conclusion}.

\section{System Model}\label{System Model}
\begin{figure}
\vspace{-2mm}
\begin{center}
\centerline{\includegraphics[width=8.75cm]{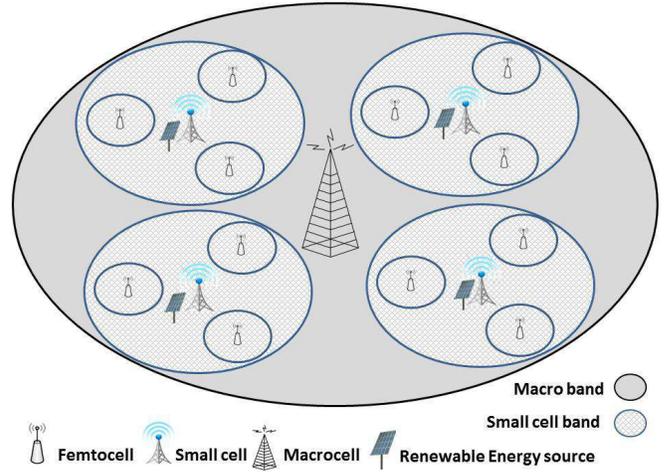}}\vspace{-0.2cm}
   \caption{\, HetNet with one macrocell BS, 4 small cell BSs, and 3 FAPs per small cell.}\label{system_model}
\end{center}\vspace{-1cm}
\end{figure}

We consider an OFDMA cellular HetNet that consists of one macrocell BS placed at the center of the cell, and $L_{s}$\footnote{For the readers' convenience, the symbol notations in this paper are summarized in Table~\ref{tabApp} of the appendix.} small cell BSs placed around the center. We assume that $L_f$ FAPs, belonging to private owners other than the mobile operator, are placed within each small cell to serve registered subscribers as shown in Fig.~\ref{system_model}. The FAPs, that are placed outside the range of small cell BSs, are not considered in the system model since they will not find a small cell BS to cooperate with in the hybrid access scenario and will not provoke interference to any of the deployed small cell BSs. In fact, in OFDMA HetNets, the available spectrum of bandwidth $B_\text{C}$ is divided into $N_{\text{C}}$ carriers. These carriers are separated depending on the type of the BS. We denote by $N_{\mathrm{C}}^{(\chi)}$, the number of available carriers at the BS of type $\chi$, where $\chi \in \{\text{M,S,F}\}$ referring to macrocell, small cell, and femtocell, respectively. It is worth mentioning that separating the available carriers into two or more groups is widely used in literature and is particularly advisable in the context of HetNets. Infact, it is recommended that small cell BSs and FAPs use higher frequency bands that are not well suited for macrocell BSs to avoid harmful interference and enhance performance~\cite{Qualcomm, 6933333}. We denote by $U$ the total number of outdoor users in the network, and $V_{il}$ the total number of indoor subscribers connected to FAP $F_{il}$, i.e., the $l^{\text{th}}$ FAP under the coverage of small cell BS $S_i$. Finally, we make the following two assumptions:
\begin{itemize}
\item A user is served by only one BS (macrocell, small cell, or FAP).
\item Each carrier is exclusively allocated to only one user.
\item There is no intra-cell interference on the downlink and no interference between macrocell BS and small cell BSs as they are using different sets of orthogonal carriers.
\end{itemize}

\subsection{Path loss and Channel Model}\label{Channel_model}
In a cellular HetNet, there are several possible radio propagation paths in the system emanating from the location of any BS and the location of the user connected to that BS. A BS can either be located outside the building (e.g., macrocell BS and small cell BSs), or inside the building (e.g., FAP). Similarly, the user can also be located either inside or outside the building. Hence, the possible propagation paths between a user and a BS are: indoor-indoor,  outdoor-indoor (equivalent to indoor-outdoor), and outdoor-outdoor. The path loss between an indoor user $v$ and a FAP $F_{il}$ (indoor-indoor path loss) is given by~\cite{Qualcomm_D2Dchannels_3GPPmeetings_MaltaFeb2013}:
\begin{align}\label{Path_Loss__indoor}
  &\mathrm{PL}_{v,F_{il},\mathrm{dB}} ^{(\text{in,in})}= 38.46 + 20\log_{10}{d_{v,F_{il}}}+0.3d_{{v},F_{il}}, \\
	&\hspace{2cm}\: \: i=1,\ldots,L_{s}, l=1,\ldots,L_{f},\nonumber
\end{align}
where $d_{v,F_{il}}$ is the distance between the indoor user $v$ and FAP $F_{il}$. In~(\ref{Path_Loss__indoor}), the first two terms, i.e., $38.46 + 20\log_{10}d_{v,{F_{il}}}$, account for the distance dependent free space path loss. The last term $0.3d_{v,{F_{il}}}$ models the indoor distance dependent attenuation.

The path loss between an outside user $u$ and a FAP $F_{il}$ (outdoor-indoor path loss) is given as follows~\cite{Qualcomm_D2Dchannels_3GPPmeetings_MaltaFeb2013}:
\begin{align}
\label{Path_Loss_outdoor_indoor}
&\mathrm{PL}_{u,F_{il},\mathrm{dB}}^{(\text{out,in})} =15.3 + 37.6\log_{10}d_{\mathrm{out}, u,F_{il}}+0.3d_{\mathrm{in},u, F_{il}} + L_{\mathrm{ow}},\\
&\hspace{2cm}\: \: i=1,\ldots,L_{s},l=1,\ldots,L_{f},\nonumber
\end{align}
where $d_{\mathrm{out}, u, F_{il}}$ is the distance between the outdoor user $u$ and the building external wall, $d_{\mathrm{in}, u, F_{il}}$ is the indoor distance between the building wall and FAP $F_{il}$, and $L_{\mathrm{ow}}$ is an outdoor-indoor penetration loss (loss incurred by the outdoor signal to penetrate the building).

The path loss between outside user $u$ and an outdoor macrocell BS or small cell BS (outdoor-outdoor path loss) is given as follows~\cite{Qualcomm_D2Dchannels_3GPPmeetings_MaltaFeb2013}:
\begin{equation}\label{Path_Loss_outdoor_outdoor}
\mathrm{PL}_{{u},\chi_{i},\mathrm{dB}}^{(\text{out,out})}= \kappa + \nu \log_{10}d_{u,\chi_{i}}, \:\: \chi_{i}=M,S_1,\cdots,S_{L_s},
\end{equation}
where $d_{u,\chi_{i}}$ is the outdoor distance between the outside user $u$ and the $i^{th}$ BS of type $\chi$. Note that for outdoor BSs, $\chi$ can only take the values in the set $\{M,S\}$ corresponding to the macrocell BS and small cell BSs, respectively. In \eqref{Path_Loss_outdoor_outdoor}, $\kappa$ and $\nu$ correspond to the path loss constant and path loss exponent, respectively.

Taking into account fading and shadowing fluctuations in addition to the path loss, the channel gain between a user $t$ ($t=u$ for outdoor user and $t=v$ for indoor user), and a transceiver station $\chi_{i}$ (BS or FAP) over a generic carrier $r$ can be expressed as:
\begin{equation}
\label{Path_Gain}
   h_{t,\chi_{i},r,\mathrm{dB}}^{(\text{x,y})} = - \mathrm{PL}_{t,\chi_{i},\mathrm{dB}}^{(\text{x,y})} + \xi_{t,\chi} + 10\log_{10}\mathrm{F}_{t,\chi_{i},r},
\end{equation}
where the first term captures the propagation loss corresponding to the considered link where $(\text{x,y}\in\{\text{out},\text{in}\})$, according to the path loss expressions given~(\ref{Path_Loss__indoor})-(\ref{Path_Loss_outdoor_outdoor}), the second term $\xi_{t,\chi}$, captures the effect of shadowing, and the last term $\mathrm{F}_{t,\chi_{i},r}$ accounts for the fast fading between user $t$ and the BS $\chi_{i}$ over carrier $r$.

\subsection{Base Station Power Model}
\label{powerModel}
Since we utilize the BS ON/OFF switching strategy in the developed framework, we consider that each BS is either in \emph{active} or \emph{sleep} mode. In the active mode, the BS is serving a certain number of users connected to the network. The power consumption of the $i^{th}$ BS of type $\chi$, denoted by $P_{\chi_{i}}$, in this mode is computed as follows \cite{EARTH}:

\begin{equation}
\label{Earth_model}
P_{\chi_{i}} = \frac{\frac{P_{\chi_{i}}^{\mathrm{tx}}}{\eta_{\text{PA}}(1-\sigma_{\text{feed}})} + P_{\text{RF}} + P_{\text{BB}}}{(1-\sigma_{\text{DC}})(1 - \sigma_{\text{MS}})(1 - \sigma_{\text{cool}})},
\end{equation}
where $P_{\chi_{i}}^{\mathrm{tx}}$ represents the transmit power radiated by BS $\chi_{i}$, $\eta_{\textrm{PA}}$ is the efficiency of the power amplifier, $P_{\textrm{RF}}$ represents the small-signal RF transceiver power, and $P_{\textrm{BB}}$ is the baseband power. The parameters $\sigma_{\textrm{feed}}$, $\sigma_{\textrm{DC}}$, $\sigma_{\textrm{MS}}$, and $\sigma_{\textrm{cool}}$ account for the feeder losses, DC-DC power supply losses, main supply losses, and cooling losses, respectively. The BS radiated power $P_{\chi_{i}}^{\mathrm{tx}}$ can be expressed as:
\begin{equation}
P_{\chi_{i}}^{\mathrm{tx}}=\sum_{r=1}^{N^{(\chi_{i})}_{\mathrm{C}}} P_{\chi_{i},r}.
\end{equation}
It corresponds to the sum of the radiated power level $P_{\chi_{i},r}$ over the carriers $r=1,\cdots, N^{(\chi_{i})}_{\mathrm{C}}$. If carrier $r$ is allocated to a certain user, then $P_{\chi_{i},r}>0$, , $P_{\chi_{i},r}=0$ otherwise.

In the sleep mode, the BSs consume power equal to $P_{\chi}^{\text{sleep}}$. The sleep mode is a reduced power consumption state in which the BS in not completely turned off and can be readily activated. Although the BS is not radiating power in this mode, elements such as power supply, baseband digital signal processing, and cooling are still active. Therefore, the BS keeps consuming power unless it is in a state of complete shutdown. For the sake of simplicity, the total power consumption of a BS given in \eqref{Earth_model} can be approximated by a linear model as follows~\cite{EARTH}:
\begin{equation}
P_{\chi_{i}} \approx \left\{
	\begin{array}{ll}
		a_{\chi} P_{\chi_{i}}^{\mathrm{tx}}+b_{\chi},  & \mbox{if } 0 < P_{\chi_{i}}^{\mathrm{tx}} \leq P_{\chi}^{\text{max}}, \\
		P_{\chi}^{\text{sleep}}, & \mbox{if } P_{\chi_{i}}^{\mathrm{tx}} = 0,
	\end{array}
\right.
\label{equationpowermodel}
\end{equation}
where $a_{\chi}$ is a factor that scales with the radiated power due to amplifier and feeder losses and $b_{\chi}$ models an offset of site power which is consumed independently of the average transmit power. We consider that the total power consumption of any BS is limited by a peak power constraint, $P_{\chi_{i}}^{\mathrm{tx}} \leq P_{\chi}^{\text{max}}$, where $P_{\chi}^{\text{max}}$ corresponds to the maximum transmit power consumption of a BS of type $\chi$ at full load. Note that $a_{\chi}$, $b_{\chi}$, and $P_{\chi}^{\mathrm{max}}$ are different for each type of BS $\chi$. Additionally, in the case of FAPs, the index $\chi_i$ is replaced by $F_{il}$, e.g., $P_{F_{il}}$ denotes the power consumption of FAP $F_{il}$.

We also consider that small cell BSs can be powered by locally generated renewable energy sources, e.g., solar, wind, etc. Each small cell BS has an available amount of renewable energy, denoted by $\bar{Q}_{i}, \: i =1,\ldots, L_{s}$, that is used to partly cover the power consumption of the small cell BSs. The availability of renewable energy is modeled as a random variable with mean $\mu_{g}$ and standard deviation $\sigma_{g}$. It should be noted that it is possible to consider that macrocell BS and FAPs are powered with renewable energy sources. However, in this study as it will be shown in Section~\ref{CoopAgr}, renewable energy is employed as a potential way for cooperation between small cell BSs and FAPs. In addition to the roaming prices, small cell BSs can share their surpluses of renewable energy with FAPs belonging to private owners to compensate the extra energy cost paid by FAPs. Considering renewable energy for the macrocell BS and FAPs will not affect the cooperation process in our context.

\subsection{Downlink Data Rates}\label{Data_rate}
The achievable data rate of the $u^\text{th}$ outdoor user served by a BS over the $r^\text{th}$ carrier, with an allocated power $P_{\chi_{i},r}$,  can be evaluated as:
\begin{equation}\label{rate_MSF}
  R_{u, \chi_{i}, r}(P_{\chi_{i},r})=\frac{B_{\text{C}}}{N_\text{C}} \log_2\left(1+\frac{P_{\chi_{i},r}\,h_{u,\chi_{i},r}^{(\text{out,y})}}{\mathcal{I}+\mathcal{N}_0}\right),
\end{equation}
where $\frac{B_{\text{C}}}{N_\text{C}}$ is the bandwidth per carrier, $\mathcal{N}_0$ is the noise power, $\text{y}$ depends on the serving BS (macrocell, small cell, or FAP), and $\mathcal{I}$ is the total interference that can be expressed as:
\begin{equation}
   \mathcal{I} = \mathcal{I}_{0} + \mathcal{I}_{u,r},
\end{equation}
where $\mathcal{I}_{0}$ is the average interference caused by neighboring small cell BSs. $\mathcal{I}_{u,r}$ is the cross-tier interference on carrier $r$ measured at the receiver $u$ caused by the closest FAPs (no intra-cell interference on the downlink direction) and expressed as follows:
\begin{equation}\label{inteferecne}
  \mathcal{I}_{u,r}=\displaystyle\sum_{i=1}^{L_s}\displaystyle\sum_{l=1}^{L_f}\left(\sum_{v_{il}=1}^{V_{il}}\upsilon_{v_{il}, F_{il}, r}\right)\cdot P_{\chi_i,r}h_{u,F_{il},r}^{(out,in)},
\end{equation}
where, $\upsilon_{v_{il},F_{il},r}$ is a binary variable representing the exclusivity of the FAP carrier allocation: $\upsilon_{v_{il},F_{il},r}=1$ if carrier $r$ is allocated to another user from a FAP $F_{il}$, and $\upsilon_{v_{il},F_{il},r}=0$ otherwise. In fact, since the same carrier might be allocated to an indoor FAP user and outdoor small cell BS user simultaneously, a cross-tier interference might be caused to each user. In each cell, an OFDMA carrier can be allocated to a single user at a given transmission time interval. Hence, we have:
\begin{equation}
\sum_{v_{il}=1}^{V_{il}}\upsilon_{v_{il},F_{il},r}\leq 1
\end{equation}
Similarly, an indoor user might be subject to an interference caused by a small cell BS serving an outdoor user. The same equations described above are kept for the data rate expression by considering the corresponding path loss expressions.

In practice, when switching off a BS, both directions (downlink and uplink) would need to be investigated. In this case, the proposed optimization problem can be considered after adding another constraint for the uplink direction, i.e., meet the uplink target data rate. However and for simplicity, we focus on the dowlink data rate only since, in the uplink, each user has its own power source (battery of the phone), whereas in the downlink the BS is the single source of power to be subdivided over all users, which makes the downlink direction more challenging. Hence, if the downlink constraints are satisfied, then uplink constraints are most probably satisfied.

\section{Network Scenarios and Optimization Problems}\label{ProblemFormulation}
The objective of the developed framework is to minimize the total power consumption of the network (i.e., macrocell BS and small cell BSs) while satisfying certain QoS requirements for each user. This will be performed by:
\begin{itemize}
\item Switching off redundant small cell BSs by optimizing a binary vector denoted by $\boldsymbol{\pi}=[\pi_{\text{S}_{1}},\cdots,\pi_{\text{S}_{L_s}}]^T$, where $\pi_{\text{S}_{i}}=0$ if small cell BS $\text{S}_{i}$ is turned off, otherwise, $\pi_{\text{S}_{i}}=1$, and $(.)^T$ denotes the vector transpose operation.
\item Optimizing the resource allocation procedure in terms of carrier-user assignment and power allocation. This will be done by optimizing the binary variables $\epsilon_{u,\chi_{i},r}$ that correspond to the status of each carrier $r$ of BS $\chi_{i}$, and the power variables $P_{\chi_{i},r}$ that are constrained by the maximum transmit power of each BS.
\end{itemize}
The QoS of the network is defined by the achieved data rate by each user, which has to be greater than a pre-defined data rate threshold denoted by $R_0$. In this section, we present the formulation of the optimization problems to determine the optimal resource and power allocation, along with the BS ON/OFF strategy for different HetNet scenarios depending on the availability and access scheme of FAPs. In the sequel, we identify the different network scenarios and provide the corresponding optimization problem. The investigated scenarios are: macro-plus-small cells (MS), macro-plus-small-plus-closed FAP (MSF-closed), and macro-plus-small-plus-hybrid FAP (MSF-hybrid), and are given as follows:

{\bf Macro-plus-small cell (MS):} In this situation, the users have the possibility to communicate with the small cell BSs in addition to the macrocell BS. Cross-interference is not considered due to the absence of FAPs (i.e., $\mathcal{I}_{u,r}=0, \, \forall u=1,\cdots, U$). Operator has the possibility to turn off small cell BSs during low traffic period. Corresponding users might be served by the macrocell BS only if its capacity allows. Thus, the optimization problem is formulated as follows:
	\begin{align}\label{Objective_fun}
&\underset{\boldsymbol{\pi},\boldsymbol{\epsilon}, \boldsymbol{P} \geq 0}{\text{minimize}} \quad P_{\text{tot}}= \sum\limits_{i = 1}^{L_s}\pi_{\text{S}_i} \Big(a_{\text{S}} \left[\sum\limits^{U}_{u=1}\sum\limits^{N^{\mathrm{(S)}}_{\text{C}}}_{r=1} \epsilon_{u,\text{S}_i,r} P_{\text{S}_i,r} \right]+b_{\text{S}}\Big)\nonumber\\
&+\sum\limits_{i = 1}^{L_s}\pi_{\text{S}_i} (1-\pi_{S_i})P_{\text{S}}^{\text{sleep}}+a_{\text{M}} \left[\sum\limits^{U}_{u=1}\sum\limits^{N^{(\text{M})}_{\text{C}}}_{r=1} \epsilon_{u,\text{M},r} P_{\text{M},r} \right]+b_{\text{M}},\\
&\text{subject to:}\nonumber
\end{align}
 (C1: Power budget constraint):
\vspace{-2mm}
\begin{equation}\label{peak power}
  0\leq \sum\limits^{U}_{u=1} \sum\limits^{N^{(\chi)}_{\text{C}}}_{r=1} \epsilon_{u,\chi_i,r}  P_{\chi_i,r} \leq P_{\chi}^{\text{max}}, \quad \forall \chi_i = \text{M},\text{S}_{1}, \ldots, \text{S}_{L_{s}},
 \end{equation}
(C2: User rate constraint):
\vspace{-2mm}
 \begin{equation}\label{rate_threshold}
\sum\limits_{\chi_{i} = \text{M}, \text{S}_{1}, \ldots, \text{S}_{L_{s}}} \sum\limits^{N^{(\chi)}_{\text{C}}}_{r=1} \epsilon_{u,\chi_{i},r} R_{u,\chi_{i},r}(P_{\chi_{i},r}) \geq R_0, \, \forall u=1,\ldots,U,
 \end{equation}
(C3: Carrier selection constraint):
\vspace{-2mm}
 \begin{equation}\label{carrier_selection}
  \sum \limits_{r=1}^{N^{\mathrm{(\chi)}}_{\mathrm{C}}}  \sum\limits^{U}_{u=1}\epsilon_{u,\chi_i,r} \leq 1, \quad ,  \forall \chi_i = \text{M}, \text{S}_{1}, \ldots, \text{S}_{L_{s}}.
 \end{equation}
The first constraint C1 given by \eqref{peak power} indicates that the total transmit power of each BS (i.e., marco cell ($\chi_i=$ \text{M}) or small cell BS ($\chi_i=$ \text{S}$_{i}$ for $i = 1,\cdots,L_s$) has to be lower than the peak power budget. The second constraint C2 given by \eqref{rate_threshold} is imposed to meet the QoS requirements for each user, and the last constraint C3 in \eqref{carrier_selection} indicates that a user can be served by only one carrier from one BS. Note that in the optimization problem, we are optimizing the transmit power of all BSs belonging to the network. However, we always keep the macrocell BS activated to ensure coverage and connectivity in the area when turning off the small cell BSs; in other words, $\pi_{\text{M}}=1$. We denote by $\boldsymbol{P}$ the vector containing all the power values $P_{\chi_i,r}$ in~\eqref{Objective_fun}, and by $\boldsymbol{\epsilon}$ in~\eqref{epsilon_eq} the binary matrix containing all the parameters $\epsilon_{u,\chi_i,r}$ with $U$ rows and $L_s\,N_{\mathrm{C}}^{(S)}+N_{\mathrm{C}}^{(\text{M})}$ columns. The matrix $\boldsymbol{\epsilon}$ is given as follows: \vspace{-0.1cm}

\small
\begin{equation}\label{epsilon_eq}
\boldsymbol{\epsilon} =
 \begin{pmatrix}
  \epsilon_{1,\text{M},1} & \cdots & \epsilon_{1,\text{M},N_{\mathrm{C}}^{(\text{M})}} &  \epsilon_{1,\text{S}_1,1} & \cdots &  \cdots & \epsilon_{1,\text{S}_{L_s},N_{\mathrm{C}}^{(\text{S})}}\\
  \vdots  & \vdots  & \vdots & \vdots & \vdots &  \vdots & \vdots\\
  \epsilon_{U,\text{M},1} & \cdots & \epsilon_{U,\text{M},N_{\mathrm{C}}^{(\text{M})}} & \epsilon_{U,\text{S}_1,1} & \cdots &  \cdots & \epsilon_{U,\text{S}_{L_s},N_{\mathrm{C}}^{(\text{S})}}\\
 \end{pmatrix}.\\
 \end{equation}
\normalsize

{\bf Macro-plus-small-plus-closed FAPs (MSF-closed):} In this scenario, only registered users can communicate with the closed FAP, outdoor users are not allowed to exploit the FAP resources. Thus, they are subject to the interference caused by deployed FAPs (i.e., $\mathcal{I}_{u,r}\geq 0, \, \forall u=1,\cdots, U$). Hence, the same optimization problem provided in~\eqref{Objective_fun}-\eqref{carrier_selection} is adopted but with the consideration of the cross-interference in the user rate expression in constraint~\eqref{rate_threshold}. In this scheme, we assume that the carriers exploited by the FAPs are known as well as the power $P_{F_{il},r}, \forall i=1,\cdots,L_s, l=1,\cdots,L_f$. For instance, they can be determined by solving the same optimization problem given in~\eqref{Objective_fun}-\eqref{carrier_selection} on each FAP or assuming a uniform power allocation.

{\bf Macro-plus-small-plus-hybrid FAPs (MSF-hybrid):} This scenario is similar to the MSF-closed, except that outdoor users are also able to exploit available FAP resources. After allocating its resource to the indoor users, a FAP can serve outdoor users by providing them with the remaining resources in terms of carriers and/or power. The optimization problem for the MSF-hybrid scenario is written as follows:
\begin{align}\label{Objective_fun1}
&\underset{\boldsymbol{\pi},\boldsymbol{\epsilon}, \boldsymbol{P} \geq 0}{\text{minimize}} \quad P_{\text{tot}}= \sum\limits_{i = 1}^{L_s}\pi_{\text{S}_i} \left(a_{\text{S}} \left[\sum\limits^{U}_{u=1}\sum\limits^{N^{\mathrm{(S)}}_{\text{C}}}_{r=1} \epsilon_{u,\text{S}_i,r} P_{\text{S}_i,r} \right]+ b_{\text{S}}\right) \nonumber\\
&+\sum\limits_{i = 1}^{L_s}\pi_{\text{S}_i}(1-\pi_{S_i})P_{\text{S}}^{\text{sleep}}+a_{\text{M}} \left[\sum\limits^{U}_{u=1}\sum\limits^{N^{(\text{M})}_{\text{C}}}_{r=1} \epsilon_{u,\text{M},r} P_{\text{M},r} \right]+b_{\text{M}},\\
&\text{subject to:}\nonumber
\end{align}
 (C1-1: Power budget constraint for mobile operator' BSs):
\vspace{-2mm}
\begin{align}\label{peak power1}
 &0\leq \sum\limits^{U}_{u=1} \sum\limits^{N^{(\chi)}_{\text{C}}}_{r=1} \epsilon_{u,\chi_i,r}  P_{\chi_i,r} \leq P_{\chi}^{\text{max}},  \nonumber\\
&\hspace{0.5cm}\quad  \forall \chi_i = \text{M},\text{S}_{i}, \text{ where } i=1,\ldots L_s,
 \end{align}
 (C1-2: Power budget constraint for FAPs):
\vspace{-2mm}
\begin{align}\label{peak powerFAP}
&0\leq \sum\limits^{U}_{u=1} \sum\limits^{N^{(\chi)}_{\text{C}}}_{r=1} \epsilon_{u,\chi_i,r}  P_{\chi_i,r} \leq P_{F_{il}}^{\text{max}},\nonumber\\
&\hspace{0.5cm} \quad \forall i=1,\ldots L_s, \text{ and } \forall l=1,\ldots,L_f,
 \end{align}
(C2: User rate constraint):
\vspace{-2mm}
 \begin{align}\label{rate_threshold1}
&\sum\limits_{\chi_{i} = \text{M}, \text{S}_{i}, F_{il}} \sum\limits^{N^{(\chi)}_{\text{C}}}_{r=1} \epsilon_{u,\chi_{i},r} R_{u,\chi_{i},r}(P_{\chi_{i},r}) \geq R_0,\nonumber\\
&\hspace{0.5cm} \quad \forall u=1,\ldots,U,\, i=1,\ldots L_s, \text{ and } l=1,\ldots,L_f,
 \end{align}
(C3: Carrier selection constraint):
\vspace{-2mm}
\begin{align}\label{carrier_selection1}
&\sum \limits_{r=1}^{N^{\mathrm{(\chi)}}_{\mathrm{C}}}  \sum\limits^{U}_{u=1}\epsilon_{u,\chi_i,r} \leq 1, \nonumber\\
&\hspace{0.2cm} \quad \forall \chi_i = \text{M},\text{S}_{i}, F_{il}, \text{ where } i=1,\ldots L_s, \text{ and } l=1,\ldots,L_f.
\end{align}
Notice that, in this case, the objective function and the rate expression given in~\eqref{Objective_fun1} and~\eqref{rate_threshold1}, respectively, remain similar to the ones of the MSF-closed scenario. However, thanks to the hybrid FAP access scheme, more degrees of freedom are offered to the mobile operator. Indeed, it has the potential to exploit the unused resources of FAPs. To enable this cooperation between the mobile operator and the FAP operators, the matrix $\boldsymbol{\epsilon}$ constructed in~\eqref{epsilon_eq} contains, in addition to the macrocell BS and small cell BSs' carriers, the remaining available FAPs' carriers. Moreover, the previous constraint C1 is divided into two constraints: C1-1 for the mobile operator BSs and C1-2 for FAPs. We notice here that the power budgets of FAPs available for the mobile operator's users, denoted by $P_{F_{il}}^{\text{max}}$, where $P_{F_{il}}^{\text{max}}\leq P_{F}^{\text{max}}$, is different for each FAP depending on the number of registered users at each FAP $F_{il}$. In this case, the problem complexity will increase compared to the other scenarios but it is expected that the MSF-hybrid scheme achieves better performances as it will be shown in Section~\ref{Performance}.

The power consumption of FAPs are ignored from the optimization problem~\eqref{Objective_fun1}-\eqref{carrier_selection1} as these FAPs are out of control of the mobile operator. Therefore, we have proposed in Section~\ref{CoopAgr} a cooperative framework for the MSF-hybrid to compensate the losses of FAPs' owners.

\section{Joint Resource allocation and ON/OFF Switching Approaches}
\label{Solutions}
The optimization problems formulated in Section~\ref{ProblemFormulation} are non-convex ones due to the existence of the binary matrix $\epsilon$ as a decision variable. In this section, we propose two approaches to solve these problems: A complex dual decomposition based method developed to find the optimum solution of the problem, and a practical but low complexity iterative algorithm that achieves sub-optimal results. In the sequel and without loss of generality, we develop the steps for each of the proposed approaches to solve the optimization problem for the MS and MSF-closed scenario. It is worth to note that the same steps can be followed to find the solutions corresponding to MSF-hybrid scenario by considering the corresponding FAPs' power budgets and the available FAPs' carriers.
\subsection{Dual Decomposition Method and Optimal Solution}
\label{Dualdecomp}
The problem in~\eqref{Objective_fun}-\eqref{carrier_selection} is satisfying the dual time sharing condition~\cite{Dual_nonconvex}. Thus, the duality gap of the non-convex resource allocation problem in OFDMA multi-carrier system is negligible as the number of carriers is sufficiently large compared to the number of users.
Hence, the dual optimization problem associated with the primal problem is given by
\begin{align}
&\underset{\boldsymbol{\lambda},\boldsymbol{\mu} \geq 0}{\text{maximize}}\quad g(\boldsymbol{\lambda},\boldsymbol{\mu}),\\
&\text{subject to:} \quad  \eqref{carrier_selection}.\nonumber
\end{align}
where $\boldsymbol{\lambda}=[\lambda_\text{M},\lambda_{\text{S}_1},\lambda_{\text{S}_1},...,\lambda_{\text{S}_{L_s}}]$ and  $\boldsymbol{\mu}=[\mu_1,\mu_2,...,\mu_{U}]$ are Lagrange vectors that contain the Lagrange multipliers associated to constraints \eqref{peak power} and \eqref{rate_threshold}, respectively.
The dual function $g(\boldsymbol{\lambda},\boldsymbol{\mu})$ is defined as follows:
\begin{align}\label{dual}
&g(\boldsymbol{\lambda},\boldsymbol{\mu})\triangleq\underset{\boldsymbol{\pi},\boldsymbol{\epsilon}, \boldsymbol{P}\geq 0}{\text{minimize}}\quad \mathcal{L}(\boldsymbol{\lambda},\boldsymbol{\mu}),\\
&\text{subject to:} \quad  \eqref{carrier_selection}.\nonumber
\end{align}
where $\mathcal{L}(\boldsymbol{\lambda},\boldsymbol{\mu})$ is the Lagrangian given as follows:
\begin{align}
\mathcal{L}&=\sum\limits_{\chi_i = \text{M}, \text{S}_{1}, \ldots, \text{S}_{L_{s}}}\pi_{\chi_i} \left(a_\chi \left[\sum\limits^{U}_{u=1}\sum\limits^{N_{\text{C}}^{(\chi)}}_{r=1} \epsilon_{u,\chi_i,r} P_{\chi_i,r}\right]+b_\chi \right)\nonumber\\
&+\sum\limits_{\chi_i = \text{S}_{1}, \ldots, \text{S}_{L_{s}}}(1-\pi_{\chi_i})P_{\chi_i}^{\text{sleep}}\nonumber\\
&\sum\limits_{\chi_i = \text{M}, \text{S}_{1}, \ldots, \text{S}_{L_{s}}} \lambda_{\chi_i} \left( \sum\limits^{U}_{u=1} \sum\limits^{N_{\text{C}}^{(\chi)}}_{r=1} \epsilon_{u,\chi_i,r} P_{\chi_i,r}-\bar{P}_{\chi}\right)\nonumber\\
&- \sum\limits^{U}_{u=1} \mu_u \left(\sum\limits_{\chi_i = \text{M}, \text{S}_{1}, \ldots, \text{S}_{L_{s}}} \sum\limits^{N_{\text{C}}^{(\chi)}}_{r=1} \epsilon_{u,\chi_i,r} R_{u,\chi_i,r}(P_{\chi_i,r})-R_0 \right).
\end{align}

Thus, by regrouping the terms including $P_{\chi_i,r}$, the dual problem in \eqref{dual} can be rewritten as follows:
\begin{align}
&g(\boldsymbol{\lambda},\boldsymbol{\mu})=\underset{\boldsymbol{\pi},\boldsymbol{\epsilon}, \boldsymbol{P}\geq 0}{\text{minimize}}\sum\limits_{\chi_i = \text{M}, \text{S}_{1}, \ldots, \text{S}_{L_{s}}}\sum\limits^{U}_{u=1}\sum\limits^{N^{(\chi)}_{\text{C}}}_{r=1} \pi_{\chi_i} \epsilon_{u,\chi_i,r} \mathcal{D}\left(P_{\chi_i,r}\right)\nonumber\\
&+ \sum\limits_{\chi_i = \text{M}, \text{S}_{1}, \ldots, \text{S}_{L_{s}}} \pi_{\chi_i} b_\chi +\sum\limits_{\chi_i = \text{S}_{1}, \ldots, \text{S}_{L_{s}}}(1-\pi_{\chi_i})P_{\chi_i}^{\text{sleep}}\nonumber\\
&- \sum\limits_{\chi_i = \text{M}, \text{S}_{1}, \ldots, \text{S}_{L_{s}}} \lambda_{\chi_i} \bar{P}_{\chi} +\sum\limits^{U}_{u=1} \mu_u R_0,\label{dual_modified}\\
&\text{subject to:} \quad  \eqref{carrier_selection},\nonumber
\end{align}
where $\mathcal{D}\left(P_{\chi_i,r}\right)=(a_\chi+\lambda_{\chi_i})P_{\chi_i,r}-\mu_u R_{u,\chi_i,r}(P_{\chi_i,r})$.
The steps to solve the dual problem can be described as follows:
 \begin{itemize}
\item {\bf Step 1:} Initialize the Lagrange multipliers values $\boldsymbol{\lambda}$ and $\boldsymbol{\mu}$.
\item {\bf Step 2:} Find the optimal value of $P_{\chi_i,r}$ for each pair $(u,r)$ by solving the following problem
\begin{equation}\label{optimal_power_min}
\underset{P_{\chi_i,r} \geq 0}{\text{minimize}}\quad \mathcal{D}\left(P_{\chi_i,r}\right).
\end{equation}
Hence, by solving \eqref{optimal_power_min}, the optimal power $P^{*}_{\chi_i,r}$ can be given as follows:
\begin{equation}\label{optimal_power}
P^{*}_{\chi_i,r}=\left[\frac{\mu_u B_\text{C}}{\ln2 N_\text{C}(a_\chi+\lambda_{\chi_i})}-\frac{\mathcal{I}+\mathcal{N}_0}{h_{u,\chi_i,r}} \right]^+,
\end{equation}
where $[x]^+=\max(0,x)$. 
\item {\bf Step 3:} Substitute the optimal power levels derived in \eqref{optimal_power} into \eqref{dual_modified}. Thus, the dual problem becomes:
\begin{align}\label{dual_with_optimal_power}
&g(\boldsymbol{\lambda},\boldsymbol{\mu})=\underset{\boldsymbol{\pi},\boldsymbol{\epsilon}\geq 0}{\text{minimize}}\sum\limits_{\chi_i = \text{M}, \text{S}_{1}, \ldots, \text{S}_{L_{s}}}\sum\limits^{U}_{u=1}\sum\limits^{N_{\text{C}}^{(\chi)}}_{r=1} \pi_{\chi_i} \epsilon_{u,\chi_i,r} \mathcal{D}\left(P_{\chi_i,r}^*\right)\nonumber\\
&+ \sum\limits_{\chi_i = \text{M}, \text{S}_{1}, \ldots, \text{S}_{L_{s}}} \pi_{\chi_i} b_\chi+\sum\limits_{\chi_i = \text{S}_{1}, \ldots, \text{S}_{L_{s}}}(1-\pi_{\chi_i})P_{\chi_i}^{\text{sleep}}\nonumber\\
&- \sum\limits_{\chi_i = \text{M}, \text{S}_{1}, \ldots, \text{S}_{L_{s}}} \lambda_{\chi_i} \bar{P}_{\chi} +\sum\limits^{U}_{u=1} \mu_u R_0,\\
&\text{subject to:} \quad  \eqref{carrier_selection}.\nonumber
\end{align}
It can be shown that \eqref{dual_with_optimal_power} is a linear assignment problem with respect to $\epsilon_{u,\chi_i,r}$ and $\pi_{\chi_i}$ and can be efficiently solved by using standard integer programming \cite{Rardin}. The solution obtained by the dual method is an asymptotically optimal solution \cite{Dual_nonconvex}.
\item {\bf Step 4:} After finding the optimal solutions $P_{\chi_i,r}^*$, $\epsilon_{u,\chi_i,r}^*$, and $\pi_{\chi_i}^*$ corresponding to the initialized Lagrange multipliers in {\bf Step 1}, we can employ the sub-gradient method to find their optimal values and thus, the optimal solution of the problem \cite{subgradient}. Hence, to obtain the solution, we can start with any initial values for the Lagrange multipliers and evaluate the optimal solutions (i.e., $P_{\chi_i,r}^*$, $\epsilon_{u,\chi_i,r}^*$, and $\pi_{\chi_i}^*$). We then update the Lagrange multipliers at the next iteration $(\tau+1)$ as follows:
    \begin{equation}
		\label{SubG1}
  \lambda_{\chi_i}^{(\tau+1)}=  \lambda_{\chi_i}^{(\tau)}-\delta^\tau \left(\bar{P}_{\chi}- \sum\limits^{U}_{u=1} \sum\limits^{N_{\text{C}}^{(\chi)}}_{r=1} \epsilon_{u,\chi_i,r} P_{\chi_i,r}\right), \, \forall \chi_i,
 \end{equation}
 \begin{equation}
		\label{SubG2}
  \mu_u^{(\tau+1)}=  \mu_u^{(\tau)}+\varpi^\tau \left(R_0-\sum\limits^{N_{\text{C}}^{(\chi)}}_{r=1} \epsilon_{u,\chi_i,r} R_{u,\chi_i,r}(P_{\chi_i,r})\right), \, \forall u,
 \end{equation}
 where $\delta^\tau$ and $\varpi^\tau$ are the updated step size according to the nonsummable diminishing step length policy (see \cite{subgradient} for more details). The updated values of the optimal solution and the Lagrange multipliers are repeated until convergence.
\end{itemize}

\subsection{Low Complexity Algorithm for Switching off Small Cell BSs}\label{Lowcomplexity}
As it will be shown in Section~\ref{Complexity_Analysis}, the run-time of the optimal dual decomposition solution proposed in Section~\ref{Dualdecomp}, where the optimization of carrier allocation, power, small cell BS ON/OFF status is simultaneously performed, is considerably high. Thus, we propose a low complexity suboptimal algorithm to cope with this issue. In this section, we propose to optimize the binary vector $\boldsymbol{\pi}$, the binary matrix $\boldsymbol{\epsilon}$, and the transmitted power vector $\boldsymbol{P}$ in an iterative way such that the algorithm complexity is significantly reduced.

The basic idea of the algorithm is to eliminate redundant small cell BSs without affecting the QoS. At each iteration, we initially consider a uniform power allocation. Then, we perform carrier allocation using standard integer programming \cite{Rardin} considering only the active BSs. This will reduce the complexity problem since the method is no more based on the sub-gradient method to find the Lagrange multipliers. In addition to that, the size of the matrix $\boldsymbol{\epsilon}$ will be only based on the carriers of the switched on BSs. At each iteration, the algorithm switches off one BS, finds the related suboptimal power and carrier allocation and verifies whether the absence of this BS degrades the QoS. If it is the case, the BS can not be eliminated. Otherwise, it can be safely switched off.

In order to solve the optimization problem formulated in \eqref{Objective_fun}-\eqref{carrier_selection} in a low complexity manner, we proceed, at each iteration, as follows:
\begin{itemize}
\item {\bf Step 1:} Simplify the optimization problem by distributing the peak power of the $\chi_i^\text{th}$ BS uniformly over its belonging carriers (i.e., $\bar{P}_{\chi_i,r}=\frac{\bar{P}_{\chi}}{N_{\mathrm{C}}^{(\chi)}}$) where $\bar{P}_{\chi_i,r}$ is the peak transmit power at the $r^\text{th}$ carrier of the $\chi_i^\text{th}$ BS. This means that constraint \eqref{peak power} can be expressed as:
\begin{equation}\label{newconst3}
P_{\chi_i,r}\leq \bar{P}_{\chi_i,r}, \forall r=1...,N_{\mathrm{C}}^{(\chi)}, \forall \chi_i=M, S_1...,S_{L_s}.
\end{equation}
Thus, the solution of the optimization problem becomes as follows:
\begin{equation}\label{Pt_subopt}
\mbox{$P_{\chi_i,r}=$}
\left\{
\begin{array}{l l}
\frac{A_0}{h_{u,\chi_i,r}},& \text{if} \quad  h_{u,\chi_i,r} \geq \frac{A_0}{\bar{P}_{\chi_i,r}},\\
0, &\text{otherwise},
\end{array}\right.
\end{equation}
where $A_0=\left(2^{\frac{R_{{0}}N_{{\text{C}}}}{B_{{\text{C}}}}}-1\right)\left(\mathcal{I}+N_0\right).$ The obtained solution derived in \eqref{Pt_subopt} means that the user $u$ served by the $\chi_i^\text{th}$ BS over carrier $r$ can achieve its data rate only if the corresponding channel is relatively good.
\item {\bf Step 2:} Compute $P_{\chi_i,r}$ for all possible (carrier, outdoor user) combinations.
\item {\bf Step 3:} Employ the combinatorial optimization approach: a standard integer programming~\cite{Rardin}, such as the Hungarian method~\cite{hungarian_method}, is used to find the best (carrier, user) combinations that maximizes the total number of served users with minimum power consumption. However, in some cases due to the modification of constraint \eqref{peak power}, uniform distribution of power may not be enough to serve all the outdoor users. Indeed, after allocating the carriers, some users may not achieve the required rate because of the power limitation as expressed in \eqref{newconst3}.
\item {\bf Step 4:} If the number of served users denoted by $U_{\text{served}}$ is less than $U$, redistribute uniformly the remaining power over the remaining carriers  ($\bar{P}_{\chi_i,r}=\bar{P}_{\chi_i,r}-\sum^{N_{\text{C}}^{(\chi)}}_{s=1}P_{\chi_i,r}$) and repeat Steps 1 to 3 for the non-served users. In fact, the peak power per carrier may increase compared to Step 1. This step is repeated until serving all users or the remaining power per carrier is not enough to achieve the user's target data rate. In fact, assuming a well-planned network, the latter case only happens if at least one of the small cell BSs is off. 
\end{itemize}
Details of the proposed low complexity algorithm are summarized in Algorithm \ref{algorithm_mod}. Once we find the BS status vector $\boldsymbol{\pi}^*$, the allocation matrix $\boldsymbol{\epsilon}^*$, and the corresponding total power consumption $\boldsymbol{P}^*$ using the low complexity suboptimal approach,  we compare the results with the optimal solution obtained using the dual decomposition method. As it will be shown Section~\ref{Simulation}, the performances of the proposed approach are close to those of the optimal solution with a notable gain in terms of computational complexity.

\begin{algorithm}[h!]
\caption{Iterative Algorithm for Green Switching off Small Cell Base Stations with Power and Carrier Allocation}
\label{algorithm_mod}
\begin{algorithmic}[1]
\small
\STATE Compute the total power consumption function $P_{\text{min}}=P_{\text{tot}}^{(0)}$ when all small cell BSs are switched on (Assume the set $\mathcal S$ contains all small cell BSs and $\pi= [1,\cdots,1]$).
\STATE Initialize for the current iteration ${\mathcal S}^{\text{iter}}={\mathcal S}$ and $L_s^{\text{iter}}=L_s$.
\REPEAT
\FOR {$\chi_i=S_1,\cdots, S_{L_s^{\text{iter}}}$}
\STATE Eliminate the small cell BS $\chi_i$ from ${\mathcal S}^{\text{iter}}, {\boldsymbol{\pi}}^{(\chi_i)}=[1\cdots 1 \underbrace{0}_{\chi_i^{\text{th}} \text{ position}} 1 \cdots 1 ]$
\STATE Initialize $\mathcal U=\{1,...,U\}$, $\mathcal N_{\text{C}}=\{1,...,N_{\text{C}}^{(S^{\text{iter}})}\}$, $U_\text{served}=0$.
\REPEAT
\STATE Compute the transmit power levels ${P}_{\chi_i,r}$ as it is given in (\ref{Pt_subopt}) for each $(u,r)\in (\mathcal U, \mathcal  N_{\text{C}})$ pairs.
\STATE Find ($u^*$, $r^*$) combinations by employing the standard integer programming in order to serve the maximum number of users with minimum power consumption.
\STATE Mark ($u^*$, $r^*$) combinations as occupied (i.e., update $\boldsymbol{\epsilon}$).
\STATE $\mathcal U=\mathcal U \setminus \{u^*\text{'s}\}$, $\mathcal N_{\text{C}}=\mathcal N_{\text{C}} \setminus \{r^*\text{'s}\}$ and $U_\text{served}=U-|\mathcal U|$.
\STATE $\bar{P}_\chi=\bar{P}_\chi-\sum\limits_{r\in \mathcal N_{\text{C}}^{(\chi)}}P_{\chi_i,r}$.
\UNTIL ($U_\text{served}=U$ $||$ $\bar{P}_{\chi}$ remains constant).
\ENDFOR
\STATE Find small cell BS $\chi_i^{\text{op}}$ that, when eliminated, provides the lowest total power consumption while satisfying the network QoS $\left(P_{\text{new}}^{(\chi_i^{\text{op}})}=\underset{\chi_i}\min \,P_{\text{tot}}^{(\chi_i)}\right)$.
\IF {$P_{\text{new}}^{(\chi_i^{\text{op}})}\leq P_{\text{min}}$}
\STATE BS $\chi_i^{\text{op}}$ is eliminated, ${\mathcal S}^{\mathrm{iter}}={\mathcal S}^{\mathrm{iter}}\setminus \{\chi_i^{\mathrm{op}}\}$, $L_s^{\mathrm{iter}}=L_s^{\mathrm{iter}}-1$ and $P_{\mathrm{min}}=P_{\mathrm{new}}^{\chi_i^{\mathrm{op}}}$.
\ENDIF
\UNTIL (No BSs can be eliminated).
\STATE The final optimal set of active small cell BSs is ${\mathcal S}^{\mathrm{iter}}$.
\normalsize
\end{algorithmic}
\end{algorithm}

\section{Cooperation Agreement for the Hybrid Access Control Scenario}
\label{CoopAgr}
In this section, we discuss the particularly interesting case of cooperation between mobile operator and the FAP operators corresponding to the MSF-hybrid scenario. In this case, private FAPs allow their resources to be used by the outdoor users of the mobile operator. The latter can avail the opportunity to turn off its small cell BSs and save considerable energy, while maintaining connectivity for its users via the host FAP. Hence, the FAP operators cooperate with the mobile operator in helping them reduce energy consumption. However, this will lead to an increased power consumption for the FAPs and hence, cooperation will not be possible if it results in monetary loss due to increased FAPs' energy bills. Therefore, in order to enable cooperation, the mobile operator needs to compensate for the additional expenses of the FAP operator.

In this framework, we consider two forms of compensation that can be exploited by the mobile operator, either by paying directly to the FAP operator or providing FAPs with cheaper renewable energy if available in excess at small cell BS sites. It is important to note that if there is a possibility to turn off a small cell BS and offload its users to the corresponding FAPs, then its locally generated renewable energy exceeding $P_{\text{S}_{i}}^{\text{sleep}}$ for small cell BS $S_{i}$ will be un-utilized. Hence, it is favorable to allow the FAPs to use this extra energy via the smart grid and cover the remaining compensation (if any) through monetary payment. Note that the mobile operator will sell the extra energy to the utility company (electricity provider) via smart grid. The FAP operator will receive an equivalent amount of electricity (to power the FAP), from the utility company at a preferential price denoted by $c_{\text{RE}}$. However, this will require mutual agreements between the mobile operator, electricity provider, and FAP operators to put in place suitable pricing and billing mechanisms.

The excess renewable energy after fulfilling the power requirements of small cell BS $i$, denoted by $q_{i}$ is defined as:
\begin{equation}\label{excess_renewable_energy}
  q_{i} = \max(0,\bar{Q}_{i} - P_{S_{i}} \Delta t ), \: \: i = 1,\ldots,L_{s},
\end{equation}
where $\Delta t$ is the BS operation time. Without loss of generality, we can assume that $\Delta t = 1$ second. The profit of the FAP operator in the closed access, i.e., un-cooperative mode, is calculated as follows:
\begin{equation}\label{profit_fap_operator}
  \Pi_{i}^{(u)} = \mathcal{R} - c_{f} \left(\sum \limits_{l=1}^{L_{f}} P_{\text{F}_{il}} \Delta t\right), \: \: i = 1,\ldots, L_{s},
\end{equation}
where $\mathcal{R}$ is the fixed revenue earned from the registered users by each FAP operator, $c_{f}$ is the cost per unit of traditional fossil fuel based energy. Under the cooperative scenario, the profit of the FAP operator can be expressed as:
\begin{align}\label{profit_fap_operator}
&\Pi_{i}^{(c)}(c_{\text{RE}},p_{r}) = \mathcal{R} - c_{\text{RE}} q_{i} + p_{r} \sum \limits_{l=1}^{L_{f}} N^{(\text{F}_{il})}_{\text{C}}\nonumber\\
&- c_{f} ( \max(0, \sum \limits_{l=1}^{L_{f}} P_{\text{F}_{il}} \Delta t  - q_{i} )) , \: \: i = 1,\ldots,L_{s},
\end{align}
where $p_{r}$ is the price charged per user to the mobile operator for offloading its users, also referred to as offloading price. Notice here that renewable energy plays a key role in the ON/OFF switching decision. Its availability may increase the chance of cooperation and hence, the mobile operator can turn off more small cell BSs. Indeed, the mobile operator will have sufficient green energy to share with FAPs which will have the possibility to reduce their fossil fuel consumption. In order to make agreements between the mobile operator and the FAP operator, the optimal values of $c_{\text{RE}}$ and $p_{r}$ must be determined. These parameters are evaluated using the following optimization framework:
\begin{align}
&\underset{p_{r}, c_{\text{RE}}}{\text{minimize}}
&&p_{r} \sum \limits_{i=1}^{L_{s}} \sum \limits_{l=1}^{L_{f}} N^{(\text{F}_{tl})}_{\text{C}}, \label{obj_energy_sharing}\\
&\text{Subject to:}\;&& \notag\\
&&&\Pi_{i}^{(c)}(c_{\text{RE}},p_{r}) \geq \Pi_{t}^{(u)}, i = 1, \ldots, L_{s}. \label{const1_energy_sharing}\\
&&& c_{\text{RE}} \leq c_{f}. \label{const2_energy_sharing}
\end{align}

The objective is to minimize the monetary payments made by the mobile operator. Instead, the focus is to use excess renewable energy to compensate the additional cost incurred by the FAP operator. The constraint in~\eqref{const1_energy_sharing} ensures that FAP operators do not lose in terms of profit as a result of cooperation. The constraint in~\eqref{const2_energy_sharing} enforces the cost of renewable energy to be lower than the cost of traditional fossil fuel energy in order to maintain the incentive for the FAP operator to buy renewable energy from the mobile operator. The problem in~\cref{obj_energy_sharing,const1_energy_sharing,const2_energy_sharing} is a convex problem and can be solved using standard linear programming methods~\cite{convex_optimization}. Explicit expressions of the optimal decision variables cannot be derived. However, there are many methods that can be employed to solve this type of problem, e.g., reducing the problem to its canonical form. In this paper, we employ the off-the-shelf solvers CVX~\cite{cvx} as the problem is convex by construction.

With the MSF-hybrid scenario, additional constraints must be imposed while turning off BSs in Algorithm~\ref{algorithm_mod} (line 15) apart from ensuring the QoS of the offloaded users. This includes respecting the power budget of the FAPs, the profitability, and price constraints given by~\eqref{const1_energy_sharing} and~\eqref{const2_energy_sharing}, respectively. In other words, the algorithm will not turn off a small cell BS if this action leads to a loss in terms of FAP operator's revenue.

\section{Simulation Results}\label{Simulation}
\subsection{Simulation Parameters}
A cellular HetNet consisting of one macrocell BS, $L_s$ small cell BSs uniformly placed around the macrocell BS, and $L_f$ FAPs per small cell BS is deployed to serve $U$ outdoor users and $L_f L_s \,V$ indoor users (we assume that the number of indoor users is the same for each FAP; $V_{tl}=V=3,\, t = 1,\ldots,L_{s}, \:l=1,\ldots,L_f$). An orthogonal OFDMA transmission is assumed where the total bandwidth of $B_{\text{C}}$ = 9 MHz is subdivided into two sets of orthogonal carriers. The first block of $6$ MHz (equivalent to $30$ carriers) is owned by the macrocell BS ($N_{\text{C}}^{(\text{M})}=30$), while the other block, of $3$ MHz (equivalent to $15$ carriers), is owned by small cell BSs and their corresponding FAPs ($N_{\text{C}}^{(\text{S})}=15)$. Hence, the same frequency blocks are reused with two consecutive small cell BSs or FAPs. We assume that all users within the small cell BSs are protected from the co-channel interference caused by other small cell BSs as they are deployed sparsely. The only interference considered in the simulations is the cross-tier interference between the small cell BS and its corresponding FAP. We consider that each user is served by a single carrier and we employ the Hungarian method~\cite{hungarian_method} for the solution of all integer programming problems in the paper. The channel parameters are selected according to the 3GPP standards~\cite{3GPP_release11a} as follows: the path loss constant $\kappa=128.1$ dB; path loss constant $\nu=3.76$; the penetration loss $L_{\mathrm{ow}} = 6$ dB; the shadowing is assumed to be log-normal with zero mean and a standard deviation $\sigma_{\xi} = 8$ dB; the fast fading between users and BSs is assumed to be Rayleigh with parameter $a$ such that $\mathbb{E}[\|a^{2}\|] = 1$. The power consumption parameters are selected according to the energy aware radio and network technologies (EARTH) model \cite{EARTH} as follows: $a_{\text{M}}=4.7$ W, $b_{\text{M}}=130$ W, $P_{\text{M}}^{\text{sleep}}= 75$ W for macrocell BS; $a_{\text{S}}=4.0$ W, $b_{\text{S}}=6.8$ W, $P_{\text{S}}^{\text{sleep}}= 4.3$ W for small cell BSs; $a_{\text{F}}=8$ W and $b_{\text{F}}=4.8$ W for FAPs. The maximum transmit power levels for the BSs are set as $P_{\text{M}}^{\text{max}} = 20$ W, $P_{\text{S}}^{\text{max}} = 2$ W, and $P_{\text{F}}^{\text{max}} = 1$ W. We assume that the energy generated from renewable sources is a Photovoltaic energy and is available at each small cell BS site. It is modeled as a Gaussian random variable with a mean of $50$ J and standard deviation of $15$ J~\cite{ElseiverPV,5697249}.

\subsection{Performance of the Proposed Resource Management Schemes}
\begin{figure}[t!]
  \centerline{\includegraphics[width=3.5in]{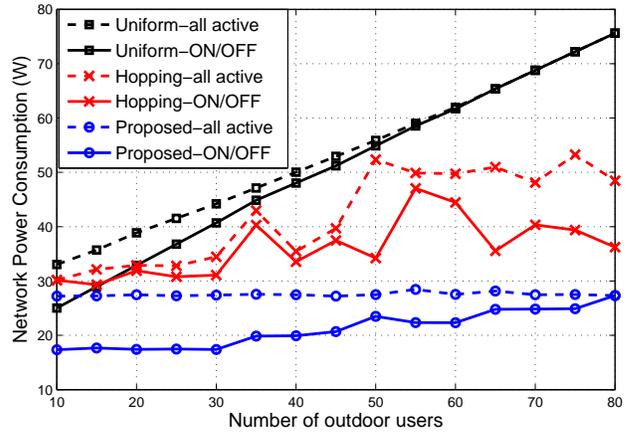}}
   \caption{\, Comparison with other resource allocation and power management schemes for $L_s=4$, $R_0=0.5$ Mbps, and values of $U$.}\label{Fig9new}
\end{figure}

We start by investigating the performances of the proposed approach, denoted by ``\emph{Proposed-ON/OFF}'', that jointly allocates the radio resources, optimizes the associated power levels, and applies the small cell BSs' ON/OFF switching scheme for the MS scenario versus the number of outdoor users $U$. In Fig.~\ref{Fig9new}, we compare its total power consumption to that of five simple combinations of the aforementioned techniques: 1) an optimized channel-aware resource allocation with uniform power management and active small cell BSs denoted by ``\emph{Uniform-all active}'', 2) same as (1 but with small cell BSs ON/OFF switching, denoted by ``\emph{Uniform-ON/OFF}'', 3) frequency-hopping resource allocation with optimized power management and active small cell BSs, denoted by ``\emph{Hopping-all active}, 4) same as 3) with small cell BSs ON/OFF switching, denoted by ``\emph{Hopping-ON/OFF}'', and 5) same as the proposed scheme but with active small cell BSs denoted by ``\emph{Proposed-all active}''. The network power consumption is defined as the sum of the power consumption of the macrocell BS, excluding $b_{\text{M}}$, and the small cell BSs. The fixed site power of the macrocell BS $b_{\text{M}}$ is a constant factor that inflates the overall power consumption and is removed to gain useful insights from the results without obscuring the differences. As it is expected, the \emph{Proposed-ON/OFF} approach outperforms the other approaches as it deals with the different parameters of the network. With \emph{Hopping-ON/OFF}, the existence of minor fluctuations in the total power consumption can be noticed, due to the randomness in the selection of the serving frequency carriers. Indeed, with the frequency-hoping scheme, the carriers are randomly allocated to a given user without taking into account the channel quality. With the \emph{Uniform-ON/OFF}, the power is linearly increasing as $U$ increases. Thanks to the ON/OFF switching important gains are obtained with all the schemes mainly for lightly-loaded networks.
\label{Performance}

\begin{figure}[t!]
  \centerline{\includegraphics[width=3.5in]{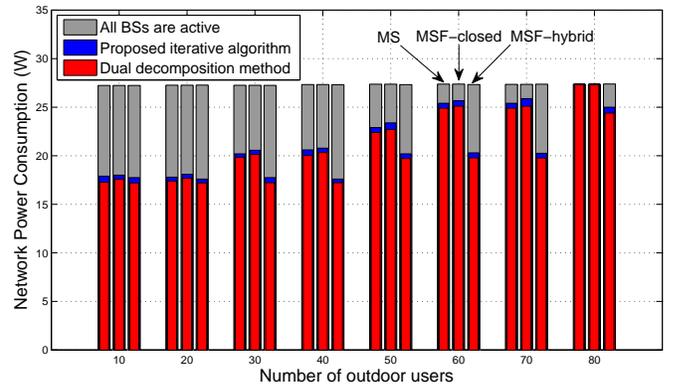}}
   \caption{\, Total network power consumption of the MS scenario versus the number of outdoor users for $L_s=4$ and $R_0=1$ Mbps.}\label{AlgorithmComparison}
\end{figure}
In Fig.~\ref{AlgorithmComparison}, we plot the total power consumption of the mobile network versus the number of outdoor users using the proposed methods, i.e., the dual decomposition method and the low complexity iterative algorithm, for three different scenarios (MS, MSF-closed and MSF-hybrid). We also compare their performances with the traditional scenario where all small cell BSs are kept active. We assume that $L_s=4$ small cell BSs are deployed and $R_0=1.0$ Mbps. In general, it can be observed that the total power consumption increases with the total number of outdoor connected users for all cases. This is because additional small cell BSs are successively activated as the number of users increase. For example, when the number of users is small, i.e., $U = 10$ or $20$, all the users are supported by the macrocell BS and the small cell BSs are inactive. However, as the number of users increase to $U=30$, one small cell BS is activated. This trend continues until all four small cell BSs are activated for $U = 80$.
The MSF-closed scenario has a slightly higher power consumption than that of the MS scenario as it considers the existence of indoor interferers that enforces BSs to increase their transmit power. On the other hand, the MSF-hybrid scenario achieves considerable energy savings, primarily due to the cooperation with the FAPs. The mobile operator is able to offload its users to the FAPs and switch off (change to sleep mode) small cell BSs to conserve energy. Compared to the traditional scenario, when all BSs are active, we can observe significant gain in terms of energy saving for the proposed scenarios. The gain is higher for small number of users, e.g., for $U = 20$, almost 40\% of energy is saved. As the number of users increases, the energy saving potential reduces since additional BSs need to be activated.


\begin{figure}[t!]
  \centerline{\includegraphics[width=3.5in]{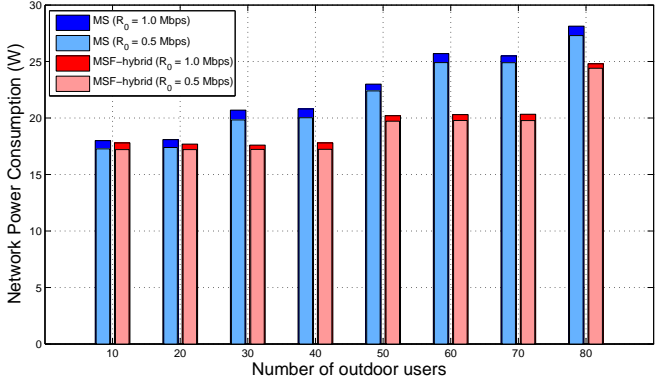}}
   \caption{\, Total power consumption using the dual-decomposition method versus the number of outdoor users for different target data rates and $L_s=4$ (MS and MSF-hybrid).}\label{versusR0_new}
\end{figure}

\indent  Fig.~\ref{versusR0_new} plots the total power consumption of the network versus the number of outdoor users in the MS and MSF-hybrid scenario for two different values of $R_0$, i.e., 0.5 and 1 Mbps. The number of deployed small cell BSs is fixed to four and $V=3$ for the MSF-hybrid scenario. The figure shows the behavior of the small cell BS ON/OFF strategy for different target data rates. We notice that the power consumption profile is the same for both scenarios and that increasing $R_0$ only imposes additional transmit power instead of activating additional small cell BSs. Note that the number of active small cell BSs can be determined from the total power consumption. Due to the dominating constant power per small cell BS $b_{\text{S}}$, we can see that the total power consumption jumps by roughly $b_{\text{S}}$ when a small cell BS is activated.

\begin{figure}[t!]
  \centerline{\includegraphics[width=3.5in]{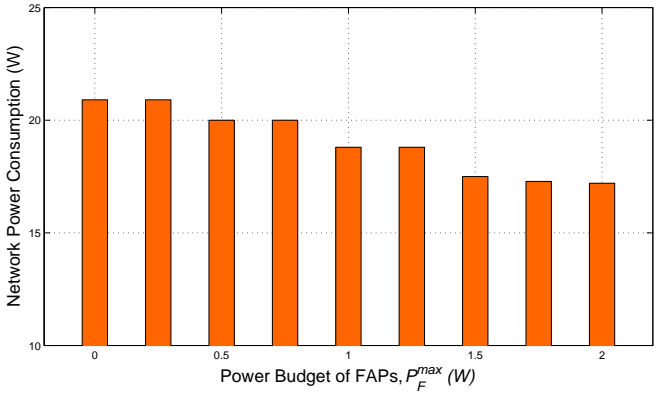}}
   \caption{\, Total power consumption using the dual-decomposition method versus the power budget of FAPs for $U=60$ and $R_0=0.5$ Mbps.}\label{versus_femtobudget}
\end{figure}

Fig.~\ref{versus_femtobudget} shows the network power consumption against the power budget, $P_{\text{F}}^{\text{max}}$ of the FAPs in the MSF-hybrid scenario for $U=60$ and $R_0=0.5$ Mbps. The transmit power of the FAPs serves as a bottleneck to support outside users. This is mainly because the channel between outdoor users and the FAPs experiences a penetration loss and thus, a higher transmit power is required to achieve the desired rate. Therefore, a FAP will only accommodate users until its transmit power capacity is reached. In Fig.~\ref{versus_femtobudget}, it can be observed that the network power consumption decreases successively as the transmit power budget of the FAPs increases. A higher power budget allows the FAPs to accept more outdoor users, thus, allowing the mobile operator to offload its users and deactivate the small cell BSs.

\begin{figure}[t!]
  \centerline{\includegraphics[width=3.5in]{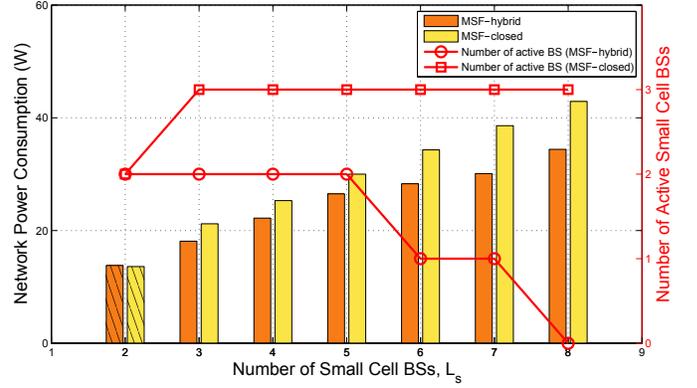}}
   \caption{\, Total network power consumption versus the number of small cell BSs.}\label{PowervsNScells_new}
\end{figure}

In Fig.~\ref{PowervsNScells_new}, we plot the total network power consumption versus the number of deployed small cell BSs $L_s$ with a constant $L_{f} = 4$. We set $U=60$ users and $R_0=0.5$ Mbps. The diagonal hatch pattern on the bars corresponding to two small cell BSs indicate that there are several users in outage due to the capacity constraints of the network. It can be noticed that having more small cell BSs provides more degrees of freedom to the network to activate the best combination small cell BSs and thus, reduce the power consumption. For the MSF-hybrid access scenario, the number of FAPs also increases with growing number of small cell BSs. Hence, there is more room for cooperation and the opportunity to deactivate small cell BSs. From Fig.~\ref{PowervsNScells_new}, it can be seen that for the MSF-hybrid scenario, the number of active small cell BSs is reduced from two to zero as the number of small cell BSs increases from two to eight. On the other hand, for the MSF-closed access scenario, the number of active small cell BSs does not change since offloading is not allowed. Nevertheless, the network power consumption increases with increasing the number of small cell BSs despite lower number of active BSs. This is because the deactivated BSs still consume $P_{\text{S}}^{\text{sleep}}$ and hence, the network power consumption increases linearly with the number of small cell BSs. Henceforth, we will only consider the MSF-hybrid access scenario since the objective is to investigate the benefits of cooperation between the mobile operator and the FAPs.

\begin{figure}[t]
\begin{subfigure}[t]{.45\textwidth}
  \centering
    \includegraphics[width=1.05\textwidth]{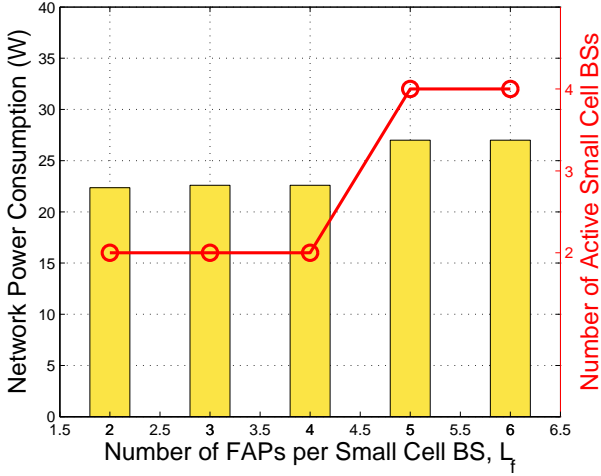}
    \caption{ }
    \label{energy_vs_smallcells}
  \end{subfigure}\hfill
  \begin{subfigure}[t]{.45\textwidth}
  \centering
    \includegraphics[width=1.05\textwidth]{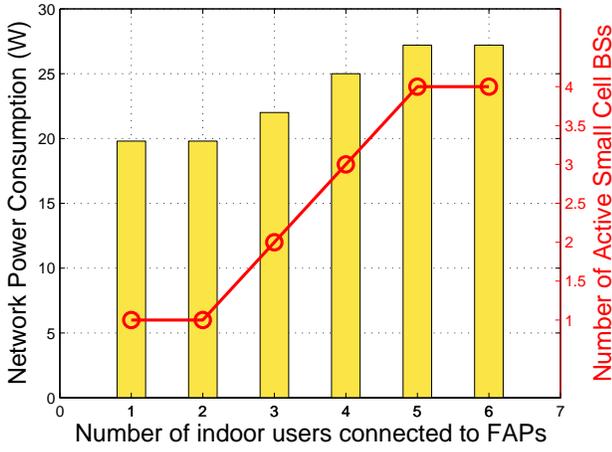}
    \caption{ }
    \label{energy_vs_faps}
  \end{subfigure}
\caption{\, Network power consumption versus (a) number of small cell BSs ($P_{\text{F}}^{\text{max}} = 0.5$ W, $R_0=0.5$ Mbps) and (b) indoor users connected to FAPs ($P_{\text{F}}^{\text{max}} = 0.5$ W, $U=60$, $R_0=0.5$ Mbps, and $L_s=4$).}
\label{energy_vs_smallcells_faps}
\end{figure}

In Fig.~\ref{energy_vs_smallcells}, the effect of varying the number of FAPs deployed per small cell BS is investigated. Increasing the FAPs leads to a higher number of registered users, which will occupy the carriers allocated to the corresponding small cell BS. Therefore, the capacity to serve outdoor users decreases and additional small cell BSs need to be activated. Fig.~\ref{energy_vs_smallcells} shows that increasing the FAPs per small cell $L_{f}$ from two to six increases the network power consumption since the number of active small cell BSs increases from two to four. On the other hand, Fig.~\ref{energy_vs_faps} investigates the network power consumption against the number of registered indoor users connected to FAPs. Similar to the behavior against number of small cell BSs, the power consumption increases as the number of indoor users increases. This is because a higher number of indoor users inhibits the FAPs from accepting outdoor users. Hence, the mobile network is not able to offload its users to the FAPs and the small cell BSs need to be activated. For example, if there is only one indoor user connected to each FAP, then only one small cell BS needs to be active and the other three are in sleep mode. The number of active small cell BSs increases with number of indoor users until all the four small cell BSs are activated.

\subsection{Behavior of the Cooperation between the Mobile and FAP Operators}
\begin{figure}[t]
  \centerline{\includegraphics[width=3.5in]{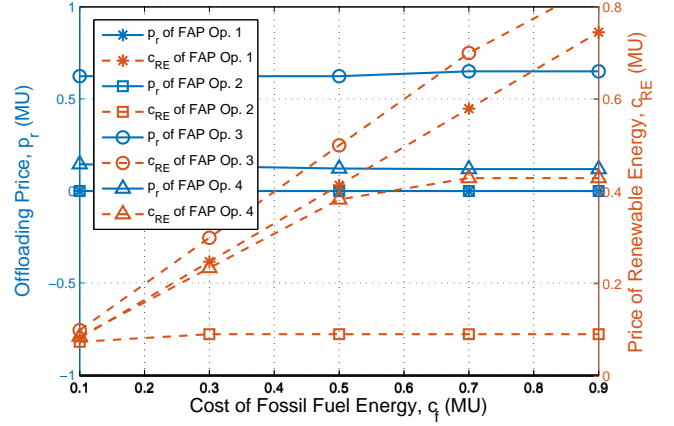}}
   \caption{\, Offloading price and price of renewable energy variation against the cost of fossil fuel energy for $\mu_{g} = 30$ J.}\label{cost_fig}
\end{figure}

\begin{figure}[t!]
  \centerline{\includegraphics[width=3.5in]{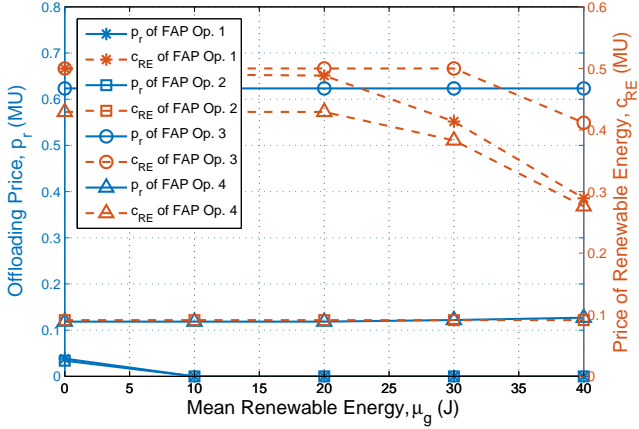}}
   \caption{\, Offloading price and price of renewable energy variation against mean renewable energy for $c_{\text{RE}} = 0.5$ MU.}\label{renewable_energy_fig}
\end{figure}

Finally, we investigate the cooperation agreements between mobile and FAP operator in the case of MSF-hybrid scenario. The cooperation agreement is based on two main parameters: 1) offloading price $p_{r}$ that serves as the monetary payment by the mobile operator to the FAP operator, and 2) the cost of renewable energy $c_{\text{RE}}$ that is purchased by the FAP operator. Fig.~\ref{cost_fig}, plots the cooperation parameters against increasing cost of fossil fuel energy. We use the same simulation settings and label the four FAP operators corresponding to the four small cell BSs as FAP Op. 1, FAP Op. 2, FAP Op. 3, and FAP Op. 4. As a result of the optimized BS ON/OFF strategy, only the small cell BS corresponding to FAP Op. 2 remains active while the rest of the small cell BSs are in sleep mode and their users are offloaded to the respective FAPs. It can be observed that FAP Op. 2 receives zero offloading price and also pays the least price of renewable energy. This is because it is not serving outdoor users and just receives the extra renewable energy available at the small cell BS at a preferential price. For the other FAP operators, it can be observed that as the cost of fossil fuel increases, the $c_{\text{RE}}$ increases in response to cover up the extra costs while the offloading price is kept to the minimum. In Fig.~\ref{renewable_energy_fig}, we plot the cooperation parameters against the mean renewable energy available at the small cell BSs. The cost of fossil fuel energy $c_{f}$ is set to $0.5$ MU. It can be observed that when the average renewable energy is zero, the offloading price for the FAP operators is non-zero, i.e., the mobile operator is required to pay all the FAP operators depending on the number of offloaded users. However, as more renewable energy becomes available, the cost of renewable energy decreases while the offloading price remains the same.

\subsection{Complexity Analysis of the Proposed Resource Allocation Schemes}
\label{Complexity_Analysis}
Regarding the employed algorithms, we can notice that the iterative algorithm is able to achieve a close performance to that of the optimal solution obtained using the dual decomposition method for all considered scenarios. The small difference is because the low complexity method does not achieve the optimal solution during the resource allocation process and the ON/OFF switching operation. Indeed, a different active small cell BSs combination in the ON/OFF switching using the iterative algorithm might lead to an increase of the energy consumption compared to the optimal combination. Despite this limitation, the iterative algorithm can considerably help in reducing the computational complexity of the problem compared to the optimal method that takes a longer time to converge. To illustrate this, we first determine the number of operations required by each approach to reach convergence and then compute the run-time (in seconds) taken by both algorithms.

For the low complexity algorithm presented in Section~\ref{Lowcomplexity}, the total number of iterations corresponding the REPEAT loop (Lines 3-19) is at maximum $\approx \left(\sum_{l=0}^{L_s-1}l=L_s\frac{(L_s-1)}{2}\right)$ in order to reach a suboptimal active small cell BS combination. At each loop of the algorithm, if we use the Hungarian method, in the worst case scenario, the maximum number of iterations for the resource allocation that can be reached is $U$ assuming that at each iteration only one pair of user and carrier is allocated (line 9 in Algorithm~\ref{algorithm_mod}). Note that, in general, we can have multiple user-carrier associations per one iteration and thus, the size of the resource allocation matrix $\boldsymbol{\epsilon}$ is reduced by more than one every iteration as indicated in line 11. Therefore, the Hungarian algorithm complexity as well as the number of operations needed to compute the power expression given in~\eqref{newconst3} depend on the size of $\boldsymbol{\epsilon}$ which is decreasing as the user-carrier associations are made and as a small cell BS is eliminated. Thus, the complexity of the low complexity algorithm is in the order of $\approx\sum\limits^{L_s-1}_{l=1}(L_s-(l-1))\left(\sum\limits^{U-1}_{u=0} \left((U-u)((L_s-l))\,N_{\mathrm{C}}^{(S)}+N_{\mathrm{C}}^{(\text{M})}-u)\right)^3\right)+\sum\limits^{U-1}_{u=0} \left((U-u)(L_s\,N_{\mathrm{C}}^{(S)}+N_{\mathrm{C}}^{(\text{M})}-u)\right)^3$ operations for the worst case scenario.

Concerning the dual decomposition method given in Section~\ref{Dualdecomp}, for a given maximum number of iterations of the sub-gradient method $I_{\text{max}}$, the algorithm is executed for $I=\text{min}(I_{\text{max}},1/\varepsilon^2)$ iterations where $\varepsilon$ represents the accuracy-guarantee which is defined by the difference between the best value and the iterate value~\cite{convex_optimization}. We can easily see that the number of operations at each iteration is in the order of $\approx I\left(\left(\left(L_s\,N_{\mathrm{C}}^{(S)}+N_{\mathrm{C}}^{(\text{M})}\right)U\right)^3+13\left(L_s\,N_{\mathrm{C}}^{(S)}+N_{\mathrm{C}}^{(\text{M})}\right)U\right)$ operations where $\left(\left(L_s\,N_{\mathrm{C}}^{(S)}+N_{\mathrm{C}}^{(\text{M})}\right)U\right)^3$ is the number of operations of the Hungarian method and  $13\left(L_s\,N_{\mathrm{C}}^{(S)}+N_{\mathrm{C}}^{(\text{M})}\right)U$ is the number of operations corresponding to the  equations given in~\eqref{optimal_power},~\eqref{SubG1}, and~\eqref{SubG2}. Note here that the optimal active small cell BSs combination is determined through the resource allocation process. By comparison, it can be observed that the iterative algorithm has a significantly lower complexity as compared to that of the optimal dual decomposition method. This is due to the fact that the terms containing $L_s$ and $U$ are successively decremented at each loop of the summation in the complexity expression of the iterative algorithm. Therefore, the resulting product is much smaller as compared to the product in the complexity expression of the dual decomposition method.

The run-time of the proposed approaches are given in Table~\ref{tabCPU}. It can be observed that the iterative algorithm requires considerably lower CPU processing time as compared to the optimal dual decomposition method. Moreover, another remark that can be deduced from Table~\ref{tabCPU} is that the MS and MSF-closed scenario requires almost the same CPU times to converge for both methods. However, the MSF-hybrid, which achieves a notable energy saving, requires more time to converge. This is because with MSF-hybrid, the size of the Hungarian matrix is larger as it also includes the redundant carriers of FAPs. All tests are performed on a desktop machine featuring an Intel Xeon CPU and running Windows $7$ Professional. The clock of the machine is $2.66$ GHz and the memory is $48$ GB. The computation time is obtained via the TIC/TOC function of Matlab.

\begin{table}[h!]
\begin{center}
\caption{\label{tabCPU} CPU times in seconds of the proposed methods}
\begin{tabular}{|c||c|c|}%
  \hline
  \textbf{} & \textbf{Iterative algorithm} & \textbf{Dual decomposition method}  \\
  \hline
	\hline
	MS   & $16.25$ & $73.53$ \\
  \hline
	MSF-closed   & $18.84$ & $76.53$ \\
  \hline
	MSF-hybrid   & $27.32$ & $118.6$ \\
  \hline
  \end{tabular}
\end{center}
\end{table}

\section{Conclusion}\label{Conclusion}
In this paper, the energy efficiency problem in dense cellular HetNets is considered. The particular case of co-existing macrocell BSs, small cell BSs, and private FAPs is investigated. Three possible network scenarios, based on the participation of private FAPs (absent, closed, hybrid), are studied. A joint strategy for radio resource and power management and BS ON/OFF switching is employed to efficiently utilize the radio access infrastructure and minimize energy consumption. The availability of locally generated renewable energy is also incorporated in the developed framework. A dual decomposition based method is proposed to achieve the optimal results. In addition, a low complexity iterative solution is proposed to achieve a near optimal solution of the non-convex problem. Results show that the cooperation between the mobile operator and the FAPs can lead to significant gains in terms of energy consumption, as compared to the non-cooperative scenario. We also developed a framework for cooperation agreements between the mobile operator and private FAPs based on monetary incentives.

The introduction of the stochastic parameters in the problem formulation can be an interesting future extension of this work. It is more practical but elaborate to consider varying user densities and varying number of FAPs per small cell BS. In addition to that, it is important to investigate the effect of the uncertainty in the channel estimations and renewable energy generation on the cooperation agreements.

\section*{Appendix : List of Parameters}
In Table~\ref{tabApp}, we summarize the most important symbols used in the document and their corresponding descriptions.
\small
\begin{table}[h]
\begin{center}
\caption{\label{tabApp} List of notations}
\addtolength{\tabcolsep}{-2.6pt}\begin{tabular}{|c|l|}%
 \hline
Symbol & Meaning  \\
 \hline
  $L_s$     & Number of small cell BSs  \\    $L_f$     & Number of FAPs \\
	 \hline
  $B_\text{C}$     & Total available bandwidth         \\\hline  $N_\text{C}$     & Number of OFDMA carriers \\
	\hline
  $N_\text{C}^{(\chi)}$ & Number of carriers allocated to BS of type $\chi$          \\\hline  $U$     &  Total number of outdoor users  \\
	\hline
 $F_{il}$     & $l^{th}$ FAP under the coverage of small cell  BS $i$  \\ \hline $V_{il}$     & Total number of indoor subscribers connected to $F_{il}$  \\
	\hline
  $\mathrm{PL}_{t,\chi_i,\mathrm{dB}}$     & Path loss between a user $t$ and BS $\chi_i$ (dB) \\\hline $d_{t,\chi_i}$ & Distance between a user $t$ and a BS $\chi_i$ \\
\hline
  $h_{t,\chi_{i},r,\mathrm{dB}}$ &  Channel between a user $t$ and BS $\chi_i$ (dB)     \\\hline  $R_{u, \chi_{i}, r}$ &   Data rate of a user $u$ served by BS $\chi_i$ over carrier $r$ \\
	\hline
   $P_{\chi_{i}}$  &  Total power consumption of BS $\chi_i$        \\\hline $P_{\chi_{i}}^{\mathrm{tx}}$ & Total transmit power of BS $\chi_i$\\
	\hline
		$P_{\chi_{i},r}$  &  Transmit power allocated over carrier $r$ of BS $\chi_i$   \\\hline $P_{\chi}^{\text{sleep}}$ & Power consumed in the sleep mode of a BS of type $\chi$ \\
	\hline
	 $a_{\chi}$  &  Coefficient scaling with $P_{\chi_{i}}^{\mathrm{tx}}$        \\\hline $b_{\chi}$ & Fixed power consumption of a BS of type $\chi$\\
	\hline
		$\mathcal I$&  Total interference      \\ \hline $R_0$ & Data rate threshold \\
	\hline
	$\pi_{S_i}$  & Status of small cell BS $S_i$\\\hline  $\epsilon_{u,X_i,r}$  & Status of carrier $r$ between user $u$ and BS $X_i$\\
		\hline
	$\mathcal L$ & Lagrangian \\\hline  $\boldsymbol{\lambda,\mu}$ & Lagrange multipliers \\
		\hline
		$\bar{Q}_i$ & Available amount of renewable energy \\ \hline $q_i$ & Excess renewable energy at small cell BS $S_i$\\
		& at small cell BS $S_i$ \\
		\hline
		$\Pi_i$ & Profit of the FAP operator in cell $i$ \\\hline $p_r$ & Cost of serving a roamed user\\
		\hline
		$c_f$ & Cost per unit of fossil fuel based energy \\\hline $c_{\text{RE}}$ & Cost per unit of renewable energy\\
		\hline
		\end{tabular}
\end{center}
\end{table}
\normalsize
\bibliographystyle{IEEEtr}
\bibliography{femto_reference}

\begin{IEEEbiography}
    [{\includegraphics[width=1in,height=1.25in,clip,keepaspectratio]{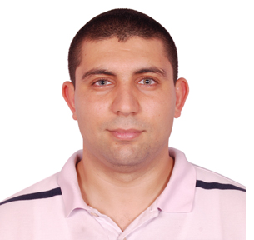}}]{Hakim Ghazzai} (S'12, M'15)
was born in Tunisia. He is currently working as a research scientist at Qatar Mobility Innovations Center (QMIC), Doha, Qatar. He received his Ph.D degree in Electrical Engineering from King Abdullah University of Science and Technology (KAUST), Saudi Arabia in 2015. He received his Diplome d'Ingenieur in telecommunication engineering and Master of Science degree from the Ecole Superieure des Communications de Tunis (SUP'COM), Tunis, Tunisia in 2010 and 2011, respectively. His general research interests include mobile and wireless networks, green communications, internet of things, UAV-based communications, and optimization.
\end{IEEEbiography}

\begin{IEEEbiography}
    [{\includegraphics[width=1in,height=1.25in,clip,keepaspectratio]{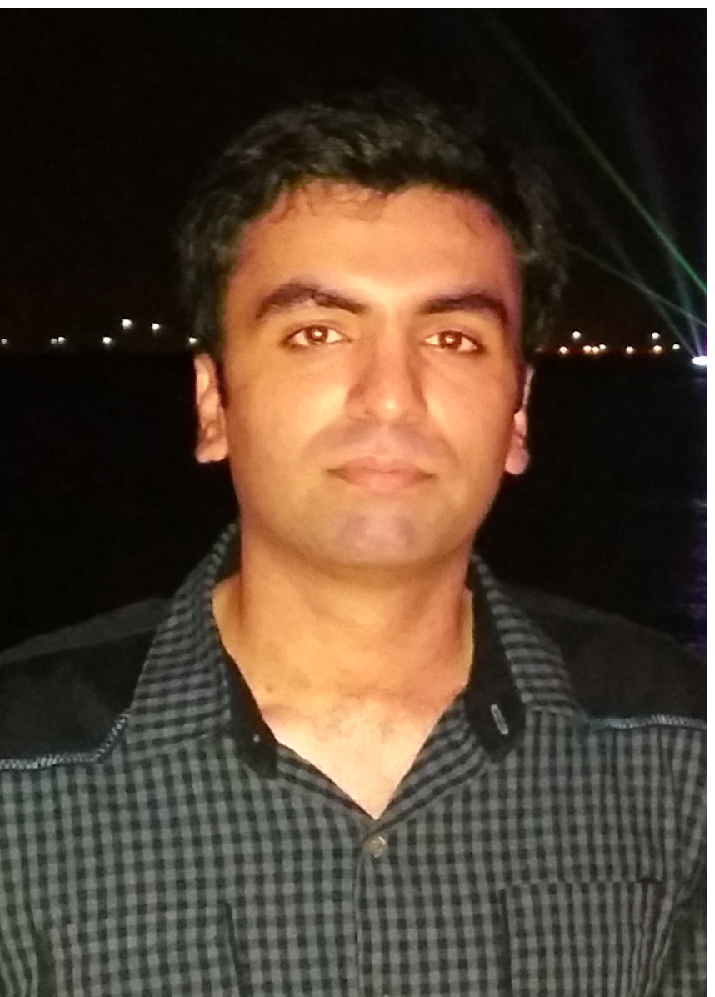}}]{Muhammed Junaid Farooq} received the B.S. degree in electrical engineering from the School of Electrical Engineering and Computer Science (SEECS), National University of Sciences and Technology (NUST), Islamabad, Pakistan, the M.S. degree in electrical engineering from the King Abdullah University of Science and Technology (KAUST), Thuwal, Saudi Arabia, in 2013 and 2015, respectively. Then, he was a Research Assistant with the Qatar Mobility Innovations Center (QMIC), Qatar Science and Technology Park (QSTP), Doha, Qatar. Currently, he is a PhD student at the Tandon School of Engineering, New York University (NYU), Brooklyn, New York. His research interests include modeling, analysis and optimization of wireless communication systems, stochastic geometry, and green communications. He was the recipient of the President's Gold Medal for the best academic performance from the National University of Sciences and Technology (NUST).
\end{IEEEbiography}

\begin{IEEEbiography}
    [{\includegraphics[width=1in,height=1.25in,clip,keepaspectratio]{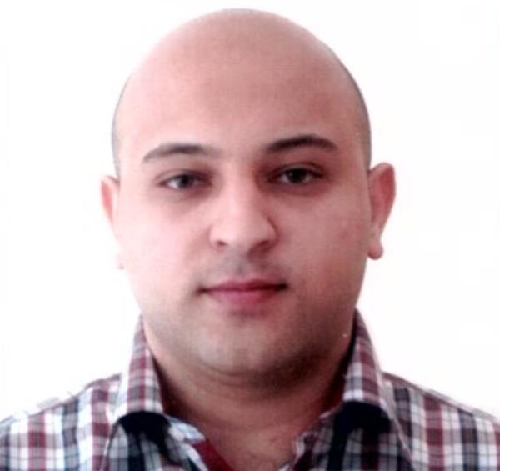}}]{Ahmad Alsharoa} (S'12) was born in Irbid, Jordan. He received the B.Sc degree (with honors) from Jordan University of Science and Technology (JUST), Irbid, Jordan, in 2011 and the M.Sc. degree from King Abdullah University of Science and Technology (KAUST) in 2013 both in Electrical Engineering. He is currently a PhD candidate in the Electrical and Computer Engineering department at Iowa State University (ISU), Ames, Iowa, USA. His current research interests include: energy harvesting, drone communications, cognitive radio networks, cooperative relay networks, MIMO communications, and green wireless communications.
\end{IEEEbiography}

\begin{IEEEbiography}
    [{\includegraphics[width=1in,height=1.25in,clip,keepaspectratio]{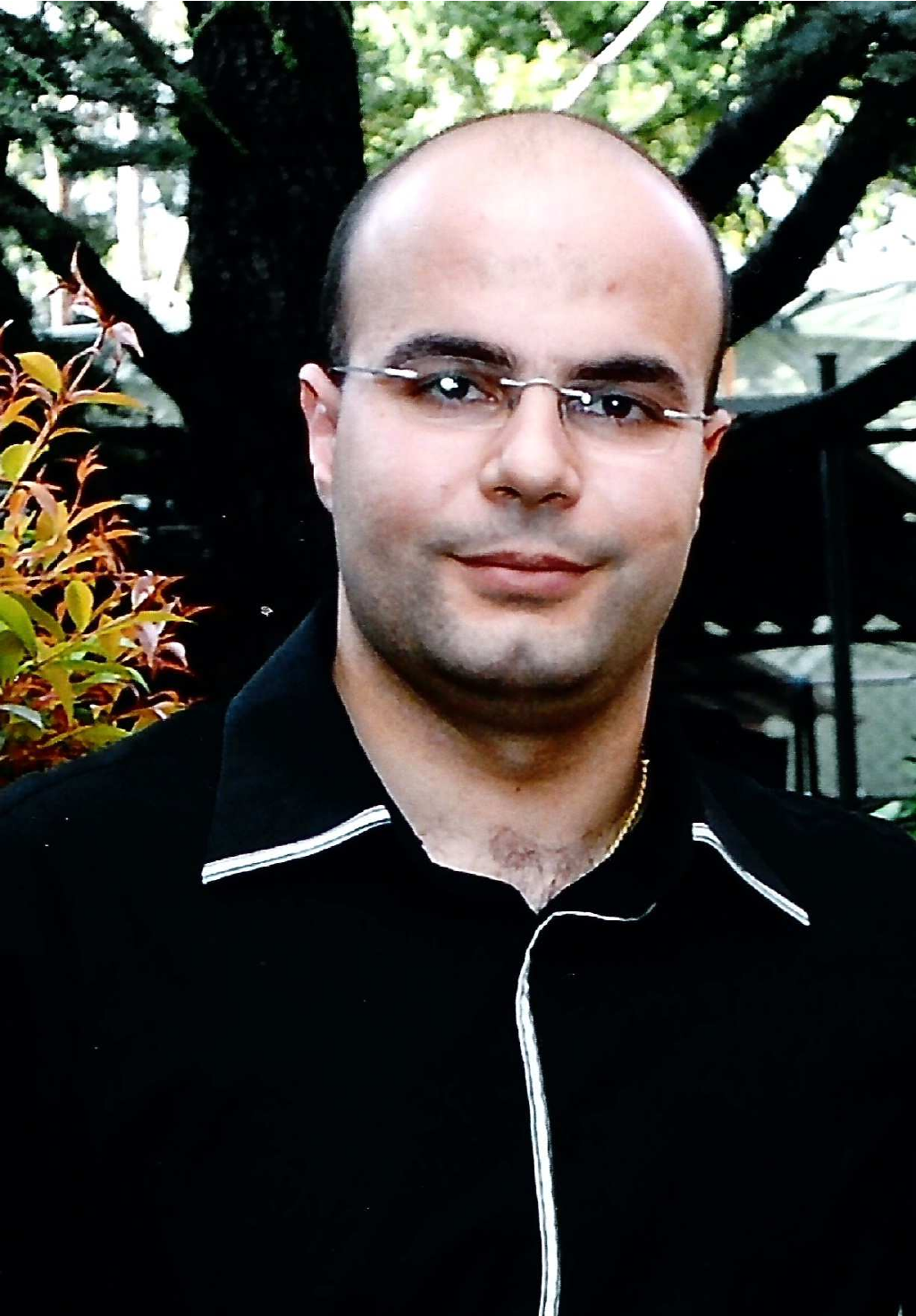}}]{Elias Yaacoub} (S'07, M'10, SM'14)
received the B.E. degree in Electrical Engineering from the Lebanese University in 2002, the M.E. degree in Computer and Communications Engineering from the American University of Beirut (AUB) in 2005, and the PhD degree in Electrical and Computer Engineering from AUB in 2010. He worked as a Research Assistant in the American University of Beirut from 2004 to 2005, and in the Munich University of Technology in Spring 2005. From 2005 to 2007, he worked as a Telecommunications Engineer with Dar Al-Handasah, Shair and Partners. From November 2010 till December 2014, he worked as a Research Scientist / R\&D Expert at the Qatar Mobility Innovations Center (QMIC). Afterwards, he joined Strategic Decisions Group (SDG) where he worked as a Consultant till February 2016. He is currently an Associate Professor at the Arab Open University (AOU). His research interests include Wireless Communications, Resource Allocation in Wireless Networks, Intercell Interference Mitigation Techniques, Antenna Theory, Sensor Networks, and Physical Layer Security.
\end{IEEEbiography}

\begin{IEEEbiography}
    [{\includegraphics[width=1in,height=1.25in,clip,keepaspectratio]{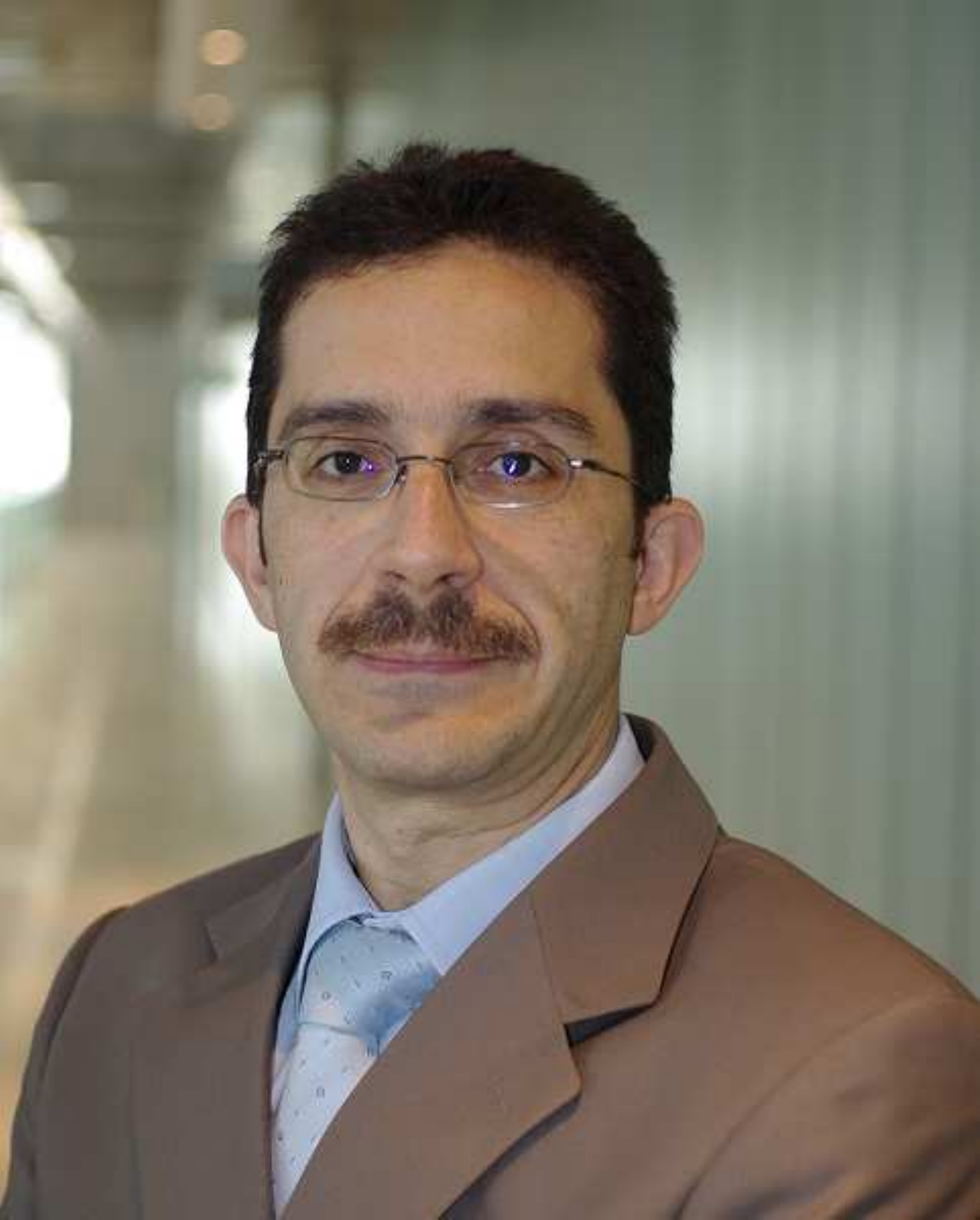}}]{Abdullah Kadri} (SM'16) received the M.E.Sc. and Ph.D. degrees in electrical engineering from the University of Western Ontario (UWO), London, ON, Canada, in 2005 and 2009, respectively. Between 2009 and 2012, he worked as a Research Scientist at Qatar Mobility Innovations Center (QMIC), Qatar University. In 2013, he became a Senior R$\&$D Expert and the Technology Lead at QMIC focusing on R$\&$D activities related to intelligent sensing and monitoring using mobility sensing. His research interests include wireless communications, wireless sensor networks for harsh environment applications, indoor localization, internet-of-things, and smart sensing. He is the recipient of the Best Paper Award at the WCNC Conference in 2014.
\end{IEEEbiography}

\begin{IEEEbiography}
[{\includegraphics[width=1in,height=1.25in,clip,keepaspectratio]{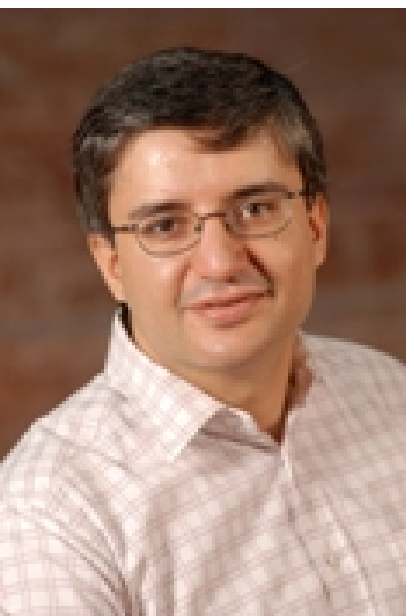}}]{Mohamed-Slim Alouini} (S'94, M'98, SM'03, F'09) was born in Tunis, Tunisia. He received the Ph.D. degree in Electrical Engineering from the California Institute of Technology (Caltech), Pasadena, CA, USA, in 1998. He served as a faculty member in the University of Minnesota, Minneapolis, MN, USA, then in the Texas A$\&$M University at Qatar, Education City, Doha, Qatar before joining King Abdullah University of Science and Technology (KAUST), Thuwal, Makkah Province, Saudi Arabia as a Professor of Electrical Engineering in 2009. His current research interests include the modeling, design, and performance analysis of wireless communication systems.
\end{IEEEbiography}

\end{document}